\documentclass[aps,prb,reprint,superscriptaddress]{revtex4-2}

\usepackage{braket}
\usepackage{amsmath, amsfonts, amssymb}
\usepackage{graphicx}
\usepackage{textcomp}
\usepackage{gensymb}
\usepackage{booktabs}

\usepackage[caption=false]{subfig}
\newcommand{\phantomsubfloat}[1]{
    {% apply caption setup only temporarily
        \captionsetup[subfigure]{labelformat=empty}
        \subfloat[][]{#1}
    }%
}

\usepackage{times}
\usepackage{verbatim}

\usepackage[bbgreekl]{mathbbol}
\DeclareSymbolFontAlphabet{\mathbbm}{bbold}
\DeclareSymbolFontAlphabet{\mathbb}{AMSb}%

\usepackage[colorlinks=true,
            linkcolor=blue,
            anchorcolor=red,
            citecolor=red]{hyperref}
\usepackage{hypernat} 
\usepackage[capitalize]{cleveref}

\begin{document}

%\preprint{}

\title{Surface states influence in the conductance spectra of Co adsorbed on Cu(111) }


\author{M. S. Tacca}
%\email[]{marcos.tacca@santafe-conicet.gov.ar}
%\thanks{}
\affiliation{Institute of Electrochemistry, Ulm University, Albert-Einstein-Allee 47, D-89081 Ulm, Germany.}
\affiliation{Instituto de F\'isica del Litoral (CONICET-UNL), G\"uemes 3450, 
S3000GLN Santa Fe, Argentina.}

\author{T. Jacob}
%\email[]{marcos.tacca@santafe-conicet.gov.ar}
%\thanks{}
\affiliation{Institute of Electrochemistry, Ulm University, Albert-Einstein-Allee 47, D-89081 Ulm, Germany.}

\author{E. C. Goldberg}
%\thanks{}
\affiliation{Instituto de F\'isica del Litoral (CONICET-UNL), G\"uemes 3450, 
S3000GLN Santa Fe, Argentina.}


\date{\today}

\begin{abstract}

We calculate the  conductance spectra of a Co atom adsorbed on Cu(111), considering the  Co
$3d$ orbitals within a correlated multiple configurations  model  interacting through the substrate band with the Co $4s$  orbital, which is  treated in a mean-field like approximation. 
By symmetry, only the  $d_{z^2}$ orbital couples with the $s$ orbital through the Cu bands, and the interference between both conduction channels introduces a zero-bias anomaly in the conductance spectra. 
We find that, while the Kondo resonance is mainly determined by the interaction of the Co $d$ orbitals with the bulk states of the Cu(111) surface, a proper description of the contribution  given by the coupling with the localized surface states to the Anderson widths is crucial to describe the interference line shape.
We find that  the coupling of the Co $4s$ orbital with the Shockley surface states is responsible of two main features observed in the  measured conductance spectra, the dip shape around the Fermi energy and the resonance structure at the surface state  low  band edge. 

\end{abstract}

%\keywords{}

\maketitle

\section{INTRODUCTION \label{sec:Introduction}}

The interaction of magnetic adatoms with the conduction electrons of the surface where they are adsorbed can lead to the formation of a Kondo resonance near the Fermi energy in the spectral density \cite{Hewson1993}. 
In scanning tunneling microscopy (STM) experiments, the Kondo resonance is usually detected as a zero-bias anomaly (ZBA) in the conductance spectra. 
The ZBA measured by STM does not always reproduce  the Kondo resonance, since interference mechanisms between the diverse conduction channels in the surface-atom-tip system can give rise to different Fano line shapes for the ZBA  \cite{Fano1961}. 
The mechanism leading to the Fano line shape has been associated with 
an interference between the correlated orbitals of the magnetic impurity that originates the Kondo resonance and further conduction channels between the tip and  the substrate \cite{Frank2015,Madhavan1998,Plihal2001,Moro-Lagares2018,Lin2006,Schiller2000,Calvo2012,Baruselli2015,Lin2005,Fernandez2021}. 
The latter channels can be due to a direct interaction of the tip and the surface \cite{Frank2015,Madhavan1998,Plihal2001,Moro-Lagares2018,Lin2006,Schiller2000,Baruselli2015,Lin2005} or indirectly via a non-interacting orbital of the impurity \cite{Frank2015,Calvo2012,Fernandez2021}. 
Alternative approaches neglect  the direct coupling of the tip with the impurity and assume  that the tunneling occurs only via tip-substrate conduction channels   \cite{Merino2004a,Merino2004,Ujsaghy2000}.

Scanning tunneling microscopy  measurements in the tunneling regime of Co atoms adsorbed on Cu show the presence of a Fano structure in the conductance spectra  \cite{Wahl2004,Neel2007,Choi2012,Knorr2002,Vitali2008,Neel2010,Limot2005}.  
Although there are several theoretical works that model  the  Co on Cu system  to different degrees of approximation  \cite{Frank2015,Baruselli2015,Merino2004a,Lin2006,Lin2005,Merino2004,Plihal2001,Fernandez2021,Dang2016}, a complete understanding of the correlated behavior of Co adatoms on Cu  requires further research \cite{Frank2015,Baruselli2015}.  
An important example is the relevance of Shockley surface states \cite{Gross2009} in the ZBA observed in  conductance spectra of magnetic  impurities adsorbed on (111) surfaces, which has been a  subject of  controversy.
On the one hand, it has been observed that the Kondo structure can be detected only with the tip within a lateral distance of $10$ \r{A} from the adatom on Au(111) \cite{Madhavan1998}, Ag(111) \cite{Schneider2002} and Cu(111) \cite{Wahl2004,Knorr2002}, suggesting a minor contribution of  the surface states to the formation of the ZBA  \cite{Plihal2001}.
The dominant contribution of the bulk states to the ZBA structures has been also supported by theoretical models using   parametric \cite{Cornaglia2003} and first-principles approaches to different degrees of approximation \cite{Barral2004,Lin2006,Merino2004a}.  
On the other hand, the quantum mirage experiment where a Kondo resonance is measured in the empty focus of an elliptical corral with a Co adatom in the other focus, shows that the contribution of surface states is certainly non negligible \cite{Manoharan2000}. 
In addition, a Co porphyrin  molecule adsorbed on a Si(111)-$\sqrt{3}\times\sqrt{3}$ Ag substrate, which does not have bulk states close to the Fermi level, evidences  the surface states contribution to the Kondo resonance in this tailored system \cite{Li2009}. 
Measurements of Co adatoms on Ag(111)  showed that the Kondo temperature  can  be tuned by  confining the surface states, 
evidencing once more their importance  \cite{Li2018,Moro-Lagares2018}.
Also with a Ag(111) substrate, terraces of different widths were used to shift  the surface state band onset above the Fermi level \cite{Henzl2007}. In this case, the  variation in the conductance spectra of Co adatoms  showed that surface states are required to observe the ZBA for lateral distances larger than $5$ \r{A}, although the Kondo temperature was determined by the bulk electrons \cite{Henzl2007}.
The latter experimental results are in agreement with theoretical predictions based on the Anderson model  \cite{Anderson1961}, in which surface states  were found to play a major role in the description of the conductance spectra of  adatoms on  metallic (111) surfaces \cite{Merino2004,Moro-Lagares2018}.
For example, it was found that the position of the surface states onset with respect to the Fermi level affects the ZBA line shape  \cite{Merino2004}.
In addition, experimental results using both magnetic and nonmagnetic metal atoms adsorbed on Cu(111) and Ag(111) surfaces showed the presence of a resonance-like feature in the conductance close to the surface state  low band edge \cite{Limot2005}. 
By means of a simple theoretical model, this structure was associated  with the coupling of the outermost orbital of the adatom with the surface states \cite{Limot2005}, which for  Co is  the $4s$. 
In a recent study \cite{Fernandez2021}, the experimental conductance of Co on Cu(111) presented in Ref. \cite{Limot2005} was used to fit the parameters of a model including the $3d_{z^2}$ and $4s$ orbitals of Co and the surface and bulk states of the Cu(111) surface. 
In Ref. \cite{Fernandez2021}, the authors find that the $4s$ orbital and its interaction with the tip play a major role in the description of the STM process.

The density of states projected on the Co $4s$ orbital, calculated by density functional theory (DFT), shows an extended  flat structure with an appreciable value around the Fermi level \cite{Baruselli2015}. 
This orbital, which is strongly hybridized with the Cu band states, is in a favorable position to interact with the tip in the tunneling regime, allowing an indirect interaction between the tip and the Cu surface \cite{Fernandez2021}.
These observations induce the proposal of considering that the interference process leading to the Fano line shape occurs on the Co adatom, between the correlated $3d$ levels and the $4s$ orbital, neglecting the direct interaction of the tip with the surface.
Although corrections related to the direct tunneling to band states may be necessary in some systems, for example to model  large lateral displacements of the tip, the approach of considering that the  interference develops between the  correlated $d$ orbitals and an essentially non-interacting level has been used to analyze conductance  in  both tunneling \cite{Frank2015} and contact  \cite{Calvo2012} regimes. 
Regarding in particular the Co on Cu(111) system that we study in this work, Ref.  \cite{Frank2015} analyze it by  considering the interference as occurring between one correlated $d$ orbital and an hybridized $sp$ orbital. 
This assumption is supported by a recent theoretical model which  showed that, among the  possible direct or indirect couplings of the tip with the Co adatom, the hopping with the Co $4s$ orbital should be dominant  in order to reproduce the  experimental findings \cite{Fernandez2021}.
Taking into account the available works on this system, such as Refs. \cite{Frank2015,Fernandez2021}, it is worth emphasizing the main contributions of our proposal. 
On the one hand, in our description we include the five $d$ orbitals in a correlated way, assuming that the non-interacting channel is provided by the Co $4s$ orbital. 
On the other hand, we calculate the required self-energies  from first-principles, and estimate   the energy levels in the same way,  remaining the absolute position of the latter as our only adjustable parameter. 
In addition, our method to calculate the Hamiltonian parameters allows a proper description of the surface states of Cu(111) and an exhaustive analysis of their influence in the  line shape of the ZBA.
We neglect the image state band, which disperses over a rather small energy range in the Cu(111) surface. 
This assumption is not directly applicable to a Cu(100) surface, where the large lifetime image state  \cite{Chulkov1999} possibly has  a key role in the interaction with the adsorbate and tip states.

For the description of the system,  we use the Anderson Hamiltonian in its ionic form \cite{Hirst1978,Hewson1993}. We extend a previously proposed multiorbital correlated model \cite{Tacca2020} to incorporate the $4s$ orbital as an additional  conduction channel treated within a mean-field like approximation that assumes no contribution to the Co spin polarization. 
We solve the ionic Hamiltonian by means of Green  functions calculated using the equation of motion (EOM) method, closing the system of equations in a second order in the atom-band coupling term \cite{Goldberg2005,Lacroix1981,Kang1995}. This method  has been used in several systems where many-body effects become relevant  \cite{Goldberg2005,Romero2009,Goldberg2017,Bonetto2016,Tacca2020,Meir1993,Feng2009}.
The introduction of the Hamiltonian and its solution are presented in \cref{sec:Theory}.

In  \cref{sec:geometry} we present a first approach to the description of the Co on Cu(111)   system, using  DFT calculations to compute the orbital occupations and estimate the total spin of the Co adatom. In \cref{sec:HamiltonianParameters} we   proceed to calculate the Anderson Hamiltonian parameters required for the description of the system with our model, namely, the Anderson self-energies and the energy levels.
One challenge in the  calculation of the conductance spectra of very dilute impurity atoms adsorbed on (111) surfaces is a first-principle calculation of the surface states contribution to the atom-surface interaction. 
Typical approaches to compute the required   Hamiltonian parameters rely on DFT supercell calculations of the surface with the impurity. 
Then, an accurate description of the surface states  requires a large number of atoms, with estimations of more than a hundred atoms per slab layer \cite{Barral2004}.
However, the number of atoms is normally restricted by the computational cost. 
In consequence, some deviations of theoretical predictions from experimental results have been associated with the necessity of improving the surface states description in the parameters calculation \cite{Baruselli2015,Frank2015}.  
Our approach for the calculation of the Hamiltonian parameters is based on a  bond-pair model \cite{Bolcatto1998} which leads to the description of the Anderson self-energies in terms of two independently calculated quantities: the dimeric couplings between the adatom and each surface atom, and the density matrix of the surface without the impurity. 
In this way, we obtain the density matrix using a primitive cell of the clean surface, avoiding supercell effects and obtaining an accurate description of the surface states. 
In addition, we are able to identify the contribution  of each band to the couplings with the different adatom orbitals and their influence in the Anderson widths.

In \cref{sec:conductance} we present  our correlated calculations of the system. 
For the conductance calculations, we model a Cu tip on top of the Co adatom and compute the couplings with the different orbitals, following the same procedure used in \cref{sec:HamiltonianParameters}. 
By assuming an hypothetical Lorentzian peak as the Kondo structure introduced by the interaction between the correlated Co $d_{z^2}$ orbital and the substrate bands, we analyze the effect of this orbital in the $s$ level. 
The interference between both channels produces a ZBA  that can be detected in the $s$ orbital spectral density.
Using our description of the Anderson widths, we identify the influence of the Cu(111) surface states in the ZBA shape.
We proceed then to present and discuss the conductance  results using our correlated model and to compare them with available data. 
Our model is able to qualitatively reproduce the experimentally observed ZBA \cite{Knorr2002,Limot2005} and the resonance-like feature close to the surface state  bands onset \cite{Limot2005}.
The conclusions are presented in  \cref{sec:Conclusions}.

\section{Theory \label{sec:Theory}}

\subsection{Ionic Hamiltonian  \label{sec:ModelExtension}}

We use the ionic Hamiltonian formalism, obtained by projecting the Anderson Hamiltonian in an adequate space of configurations  \cite{Hirst1978,Ovchinnikov2004,Hewson1993}. 
The crystal field lifts the degeneracy of the atomic configurations and  the lower  symmetry   leads to the quenching of the orbital angular momentum. 
Then, we assume that the ground state of the atom becomes an orbital singlet with angular momentum $\braket{\hat{\mathbf{L}}}=0$, so that the atomic configurations are determined by the total spin $S$  and the spin projection $M$, and are degenerated in $M$. 
We consider the infinite-U approximation in a strong Hund's rule coupling regime, in which the configuration space is restricted to states with total spin $S$ and $S-\frac{1}{2}$ \cite{Hewson1993}. 
The  infinite-U approach has been used for the description of atom-surface interacting systems in out-of-equilibrium dynamical and stationary situations \cite{Goldberg2005,Romero2009,Goldberg2017,Bonetto2016,Meir1993,Feng2009,Garcia2009,Romero2011,Bonetto2014,Tacca2017,Tacca2020}. 
In particular, in  Ref. \cite{Tacca2020} we applied it to the study of a Co adatom on graphene, including  the five Co $d$ valence orbitals in a correlated way.  
In this work we extend our correlated $d$ orbitals model to  incorporate the Co $4s$ orbital, treated within an independent electron approximation. 
For the $s$ orbital we keep the description based on fermionic operators, and we consider the same orbital energy $\epsilon_{s}$ for both spin projections within a mean-field picture.
In this way, the $s$ orbital does not contribute to the spin polarization of the adsorbed Co atom.
Working in the hole picture,  we obtain the following Hamiltonian that describes the Co adatom on the surface: 
% 
% \begin{small}
\begin{equation} \label{eq:Hionicowiths}
\begin{split}
 \hat{H}=&\sum_{\mathbf{k},\sigma}\epsilon_{\mathbf{k}}\hat{n}_{\mathbf{k}\sigma}
%     \\   &
    +\sum_{M,p}E_{S,p}\ket{S,M}_{p}\bra{S, M}_{p}
    \\   &
     +\sum_{m,q}E_{ S- \frac{1}{2} , q}\ket{S-\textstyle {\frac{1}{2}} ,m}_{q}\bra{S-\textstyle {\frac{1}{2}} ,m}_{q}
    \\ & + 
    \sum_{\mathbf{k},\sigma, M,p,q}
%     \Biggl(
    \left( 
    V_{\mathbf{k}SM\sigma}^{pq} \hat{c}_{\mathbf{k}\sigma}^{\dagger}\ket{S-\textstyle {\frac{1}{2}} ,M-\sigma}_{q}\bra{S,M}_{p}
+H.c. \right)
%     \\  &  \phantom{\sum_{\mathbf{k},\sigma M,p,q}} +H.c. \Biggr)
    \\  & 
    + \sum_{\sigma}\epsilon_{s}\hat{n}_{s\sigma} 
    + \sum_{\mathbf{k},\sigma} \left(V_{\mathbf{k}s} \hat{c}_{\mathbf{k}\sigma}^{\dagger}\hat{c}_{s\sigma}+H.c.\right).
\end{split}
\end{equation}
% \end{small}

In \cref{eq:Hionicowiths}, $\hat{n}_{\mathbf{k}\sigma}$ is the number operator for a hole with spin projection $\sigma=\pm1/2$ in the $\mathbf{k}$-band state, with energy $\epsilon_{\mathbf{k}}$. 
The terms related to the Co $d$ orbitals are written using Hubbard  projection operators $\ket{S,M}\bra{S,M}$ \cite{Ovchinnikov2004}, being $S$ ($S-\textstyle {\frac{1}{2}}$) the larger (lower) total spin and $M$ ($m$) its projection. 
The indices  $p$ and $q$ identify the  orbitals occupied by holes in each state. For example, $p=d_{xz}d_{yz}d_{z^2}$ when those three orbitals are occupied by holes in a $S=\textstyle {\frac{3}{2}}$ configuration. If the hole in the $d_{z^2}$ orbital of the latter configuration is transferred to the substrate, we obtain the $q=d_{xz}d_{yz}$ state with  $S-\textstyle {\frac{1}{2}}=1$.
The quantities $E_{S,p}$ and $E_{s,q}$ are the total energies of the configurations $\ket{S,M}_{p}$ and $\ket{S- {\frac{1}{2}},M-\sigma}_{q}$, respectively. 
The total energies define the single particle energy level active in the transition between the  configurations labeled with $p$ and $q$, $\epsilon_{d(p,q)}=E_{S,p}-E_{S- {\frac{1}{2}},q}$. 
We identify the orbital active in the transition between configurations $p$ and $q$ with $d(p,q)=d_i$. In the previous example,  $d(p=d_{xz}d_{yz}d_{z^2},q=d_{xz}d_{yz})=d_{z^2}$.
The  coupling between the substrate bands and the Co $d$ orbitals is written   using the fermionic creation operator  $\hat{c}_{\mathbf{k}\sigma}^{\dagger}$ for the $\mathbf{k}$-band states, and the  annihilation projector operator that acts in the  selected space of the Co atom, $\ket{S-\textstyle {\frac{1}{2}} ,M-\sigma}_{q}\bra{S,M}_{p}$. The coupling parameter is given by \cite{Hewson1993}
\begin{equation}\label{eq:hopping}
V_{\mathbf{k}SM\sigma}^{pq}=\bra{S-\textstyle {\frac{1}{2}},M-\sigma}_{q}\hat{c}_{d(p,q)\sigma}\ket{S,M}_{p}V_{\mathbf{k}d(p,q)} , 
\end{equation}
where $V_{\mathbf{k}d_i}$ corresponds to the coupling between the Co $d_i$ orbital and the $\mathbf{k}$-band state, and $\hat{c}_{d_i\sigma}$ is the fermionic annihilation operator that acts in the $d_i$ orbital. The  Hermitian conjugate is indicated with $H.c.$.

On the other hand, the coupling term corresponding to the Co $s$ orbital uses fermionic operators  defined in the hole picture, with a  coupling parameter $V_{\mathbf{k}s}$ between the $\mathbf{k}$-band state and the $s$ orbital.

\subsection{Green functions and conductance expressions \label{sec:Greenfunctions}}

The solution is found in terms of Green functions like the following: 
 \begin{equation}
 \label{eq:Ggen}
G_{\tilde{a}}^{\tilde{b} }(t',t)= i\theta(t'-t)\braket{ \left\{ \hat{A}^{\dagger}(t'),\hat{B}(t)\right\} } ,
\end{equation}
where $\braket{ \bullet }$ is the mean value taken on the Heisenberg representation and $\left\{\bullet,\bullet \right\}$ is the anticommutator of two operators. 
Both $\hat{A}$ and $\hat{B}$ are replaced with   either  $\ket{S,M}_{p}\bra{S-\textstyle{\frac{1}{2}},M-\sigma}_{q}$ or $\hat{c}_{s\sigma}$, and the $\tilde{a}$ and $\tilde{b}$ indices are respectively replaced with $pq$ and $ss$. In this way, we define four types of Green functions:  $G_{pq}^{pq} = G_{pq} $, $G_{pq}^{ss}$, $G_{ss}^{pq}$ and $ G_{ss}^{ss} = G_{ss} $ (see \cref{ap:Derivation}).

The Green functions are calculated by using the equations of motion method (EOM)  \cite{Lacroix1981,Kang1995} and closing the system of equations that involve an increasing number of particles in a second order in the atom-band coupling term \cite{Goldberg2005}. 
The resolution method has been  extensively discussed in different applications of this approach   \cite{Goldberg2005,Romero2009,Goldberg2017,Bonetto2016,Tacca2020,Meir1993,Feng2009}, and therefore we present the derivation and final expressions  in \cref{ap:Derivation}.

In the steady-state limit the Green functions are translationally invariant in time, so that the solution is given by the Fourier transform of the EOMs. 
The equilibrium Green functions corresponding to the $s$  ($G_s$) and each $d_i$ ($G_{d_i}$) orbitals of the adatom are then built using the Green functions defined in \cref{eq:Ggen}, 
\begin{subequations}  \label{eq:allGds} 
\begin{align} 
\label{eq:Gd} 
 G_{d_{i}}(\omega)={}&\gamma_{S}\sum_{p,q}\delta_{d(p,q)d_{i}}G_{pq}(\omega)
\\ \label{eq:Gs} 
G_{s}(\omega)={}&
2G_{ss}^{0}(\omega)+G_{s}^{(c)}(\omega) ,
\end{align}
\end{subequations}
where $\gamma_{S}=2S+1$. The mixed Green function ($G_{s}^{d_i}$, $G^{s}_{d_i}$)   are defined analogously to \cref{eq:Gd}. 
The independent particle Green function $G_{ss}^{0}$ in \cref{eq:Gs}  is   given by
\begin{equation} \label{eq:Gss0}
 G_{ss}^{0}(\omega)=\frac{1}{\omega-\epsilon_{s}-\Sigma_{s}^{0}(\omega)} .
\end{equation}
In  presence of the correlated $d$ orbitals, the $s$ Green function is modified by  $G_{s}^{(c)}$, which is given by
\begin{align} 
% \\
\label{eq:Gsc}
G_{s}^{(c)}(\omega)={}&
\sum_{d_{i}} \left(\sigma_{sd_{i}}(\omega)\right)^{2}G_{d_{i}}(\omega)
,
\end{align}
where
\begin{equation} \label{eq:sigma}
 \sigma_{sd_{i}}(\omega)=\Sigma_{sd_{i}}^{0}(\omega)G_{ss}^{0}(\omega).
\end{equation}

The Anderson self-energies introduced in \cref{eq:Gss0,eq:sigma} are given by 
\begin{equation} \label{eq:SelfEnergy}
\Sigma_{ab}^{0}(\omega)=\sum_{\mathbf{k}}\frac{V_{\mathbf{k}a}^{*}V_{\mathbf{k}b}}{\omega-\epsilon_{\mathbf{k}}-i\eta}   ,
\end{equation}
where  $a$ and $b$ are replaced by $s$ or $d_i$,   $\Sigma^{0}_{a}\equiv \Sigma^{0}_{aa}$ and $i \eta$ is an infinitesimal imaginary quantity.
In our approach, the spectral density corresponding to each orbital is given by the imaginary part of the corresponding Green function, \cref{eq:Gd} or \cref{eq:Gs}: 
\begin{equation} \label{eq:SpectralDensity}
\rho_a(\omega)=\frac{1}{\pi} \text{Im}G_{a}(\omega) .
\end{equation}

In the near-equilibrium situation, at low temperature and small bias, we calculate  the conductance $G(V)$  between the adatom and a tip using \cite{Meir1992,Calvo2012} 
\begin{equation} \label{eq:conductanceeq}
G(V)=G_0  \left( T_{d}(eV) + T_{s}(eV) + T_{sd}(eV) \right)  ,
\end{equation}
 where  the transmission is decomposed into 
\begin{subequations}
\begin{align}
\label{eq:Td}
 T_{d}(\omega)={}&\sum_{d_{i}}\Gamma_{d_{i}}^{eff}(\omega)\text{Im}G_{d_{i}}(\omega)
\\
T_{s}(\omega)={}&\Gamma_{s}^{eff}(\omega)\text{Im}G_{s}(\omega)
\label{eq:Ts}
\\
\label{eq:Tds}
T_{sd}(\omega)={}& \sum_{d_{i}}\Gamma_{sd_{i}}^{eff}(\omega) \left( \text{Im}G_{d_{i}}^{s}(\omega) + \text{Im}G_{s}^{d_{i}}(\omega) \right),
\end{align}
\end{subequations}
and  $G_0=2e^2/h$ is the quantum of conductance.
In the previous expressions, $\Gamma^{eff}$ is defined as 
\begin{equation} \label{eq:gammaeff}
 \Gamma_{ab}^{eff}(\omega)=\frac{2\Gamma_{ab}^{0}(\omega)\Gamma_{ab}^{0\text{-}tip}(\omega)}{\Gamma_{ab}^{0}(\omega)+\Gamma_{ab}^{0\text{-}tip}(\omega)} ,
\end{equation} 
where $\Gamma^0_{ab}$ and $\Gamma^{0\text{-}tip}_{ab}$ are the Anderson widths given by the imaginary parts of the  self-energies  corresponding to the interaction of the Co adatom with the surface and tip, respectively (\cref{eq:SelfEnergy}). As before, we abbreviate the notation by using $\Gamma_{a} $ instead of $ \Gamma_{aa} $. 

\section{Geometry and symmetry considerations \label{sec:geometry}}

We performed DFT calculations of the Co adatom on the Cu(111) surface to obtain the geometrical structure of the system and the orbital occupations of the adatom. 
We used the SeqQuest code  \cite{Feibelman1987,Verdozzi2002}   with the PBE \cite{PBE1996,PBE1997} functional and a force convergence criterion  of $0.01$ eV/\r{A}. 
The obtained bulk lattice parameter for Cu was $3.62$ \r{A}.
We performed $4\times4$ supercell calculations  of a (111) surface slab with five atomic layers and including the Co adatom, adding  $15$ \r{A} of vacuum to ensure the decoupling between surfaces and relaxing the system. 
We found that the preferential adsorption site is on hollow, in agreement with previous works \cite{Huang2008,Baruselli2015}. 
We found an adsorption height of the Co adatom of $1.69$ \r{A},   ${\approx}5$\% lower than the value reported in  Ref.  \cite{Huang2008} for the same system but using a different DFT code (CASTEP \cite{Segall2002}) and other  functional  (LDA \cite{LDA1981}).

The $C_{3v}$ symmetry splits the $d$ orbitals into the  groups  $E1$ ($d_{xz}$, $d_{yz}$), $E2$ ($d_{x^2-y^2}$, $d_{xy}$) and  $A1$ ($d_{z^2}$). 
The energy levels and self-energies depend only on the symmetry of the involved orbital, for example, $  \epsilon_{d_{xz}}=\epsilon_{d_{yz}} \equiv \epsilon_{E1}$. Then, a  similar notation is used for the remaining energy levels and for the self-energies.

In  \cref{tab:DFTOrbitalOccupation} we present the orbital occupations per spin for each group, obtained from DFT calculations by using  L\"odwin population analysis.
The obtained occupations for the $d$-shell suggest  fluctuations between configurations with seven and eight electrons, in agreement with previous calculations \cite{Surer2012,Huang2008}.
Then, we will consider states with total spin values of $S= \textstyle \frac{3}{2}$ and $S- \textstyle  \frac{1}{2}=1$, that is, fluctuations between configurations with three and two holes in the $d$-shell. 

\begin{table}[h] 
\begin{center}
\begin{tabular}{ccccccccc}
\toprule 
 $E1$  &  $E2$  &  $A1$  &  $d$-shell & $s$ & $p_x/p_y$ & $p_z$ & $sp$-shell   \\ % 
\midrule
 $0.77$ & $0.78$  & $0.77$ & $7.74$ & $0.25$ & $0.07$ & $0.05$  & $0.88$  \\ % 
\bottomrule
\end{tabular}
\end{center}
\caption{Orbital occupations per spin projection of Co adsorbed on Cu(111)  and total occupation of the $d$-shell and  $sp$-shell.  \label{tab:DFTOrbitalOccupation}}
\end{table}

The splitting of the Co $d$ orbitals into three groups leads to five different ways to accommodate the three holes of the $S=\frac{3}{2}$ states into the five $d$ orbitals. Five possibilities correspond  also to the two-holes configurations of the $S-\frac{1}{2}=1$ states. The transitions between these configurations lead to eleven non-equivalent fluctuations, that is, eleven different Green functions to calculate. We present the possible configurations for each spin state and the corresponding  transitions  in  \cref{ap:configurations}.

\section{Hamiltonian parameters \label{sec:HamiltonianParameters}}

\subsection{Self-energies \label{sec:SelfEnergies}}

The most relevant quantities for our calculation are the Anderson widths or hybridization functions, given by the imaginary part of \cref{eq:SelfEnergy}:
\begin{equation}
\label{eq:Gammadefinition}
\Gamma_{ab}^{0}(\epsilon)= \pi\sum_{n,\mathbf{k}}V_{n\mathbf{k}a}^{*}V_{n\mathbf{k}b}\delta(\epsilon-\epsilon_{n\mathbf{k}})
.
\end{equation}

In \cref{eq:Gammadefinition}, the indices $a$ and $b$ correspond to $s$ or $d_i$. In this section, we write explicitly the band index $n$ in the atom-band coupling terms $V_{n\mathbf{k}a}$. 

We calculate  the atom-band coupling terms $V_{n\mathbf{k}a}$  by using the bond-pair model \cite{Bolcatto1998}. The model express   $V_{n\mathbf{k}a}$ in terms of the coefficients of the density matrix of the surface without the impurity and the  dimeric couplings between the impurity and each surface atom in the symmetrically orthogonalized basis \cite{Bolcatto1998}. This expansion of $V_{n\mathbf{k}a}$ leads to the following expression \cite{Tacca2020} 
\begin{equation} \label{eq:vkas1}
V_{n\mathbf{k}a}^{*}V_{n\mathbf{k}b}= 
\sum_{\alpha,\beta, r,t}e^{i\mathbf{k}\cdot\left(\mathbf{L}_{l(r)}-\mathbf{L}_{l(t)}\right)} {c}_{\alpha h(r)}^{n\mathbf{k}}\left( {c}_{\beta h(t)}^{n\mathbf{k}}\right)^{*}V_{\alpha r a}^{*}V_{\beta tb}.
\end{equation}
In \cref{eq:vkas1}, $V_{\alpha r a}$  corresponds to the symmetrically orthogonalized coupling between the $a$ orbital of the impurity and the $\alpha$ orbital of the $r$ surface atom \cite{Bolcatto1998}. The coefficients $c_{\alpha h}^{n\mathbf{k}}$ define the symmetrically orthonormalized  density matrix  of the surface, 
\begin{equation} \label{eq:densitymatrix}
\rho_{\alpha \beta rt}(\epsilon)=\sum_{n,\mathbf{k}}e^{i\mathbf{k}\cdot\left(\mathbf{L}_{l(r)}-\mathbf{L}_{l(t)}\right)} c_{\alpha h(r)}^{n\mathbf{k}} \left(c_{\beta h(t)}^{n\mathbf{k}}\right)^{*} \delta(\epsilon_{n\mathbf{k}}-\epsilon) ,
\end{equation}
being $\mathbf{L}_{l(r)}$ the Bravais lattice vector of the $l(r)$ unit cell, where the $r$ atom is located. We identify each atom inside the unit cell with a  basis index  $h(r)$. 
We write explicitly the Bravais lattice vector  to remark that the $c_{\alpha h}^{n\mathbf{k}}$ coefficients can be calculated using a primitive cell of the surface without the impurity, which allows us to identify the couplings of the Co orbitals with each surface band. 
Using this method also allows us to determine the Anderson widths avoiding supercell effects \cite{Tacca2020}.
Note that we use the term density matrix to refer us to the object defined by \cref{eq:densitymatrix}, which are the matrix elements of the spectral function.
We should notice that the described approach is adequate for very dilute impurity atoms.

The coefficients of the density matrix of the Cu(111)  surface  were obtained from DFT calculations of a $1\times1$  slab, with nine atomic layers and a vacuum separation of $15$ \r{A}.
We used a $100\times100$ $\mathbf{k}$-point  grid for the calculation of the band structure and  density matrix. 
We verified that the  Anderson widths have converged with the number of Cu atoms by including up to nine Cu neighbors in the calculation given by  \cref{eq:vkas1}.

\begin{figure*}[ht]
\centering
\includegraphics[width=0.99\linewidth,keepaspectratio]{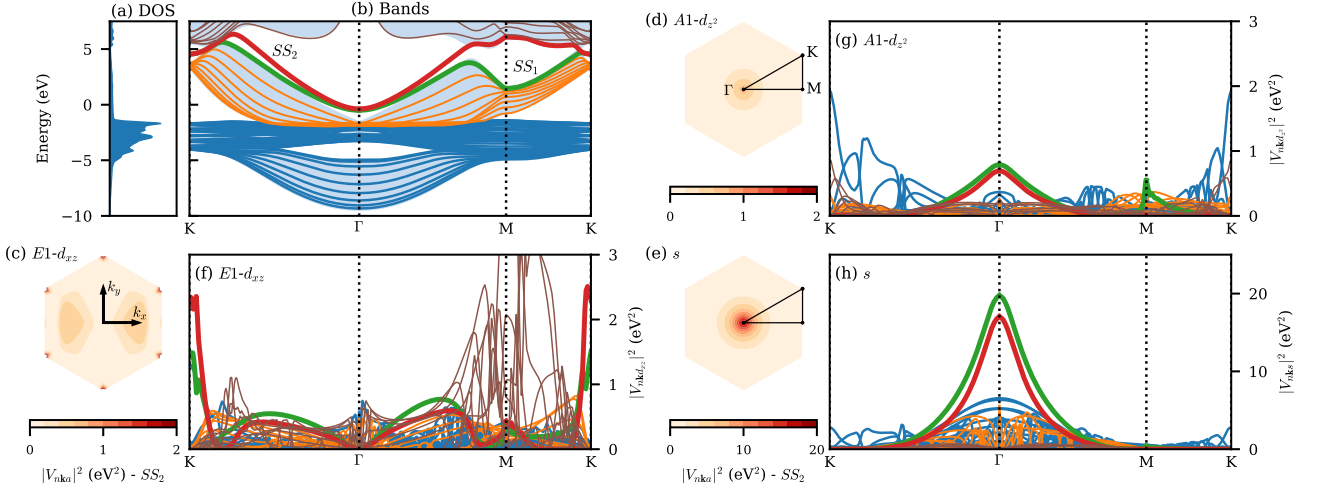}%
% Keep above the caption to avoid messing up counters
\phantomsubfloat{\label{fig:Vka_CoCu111:a}}%
\phantomsubfloat{\label{fig:Vka_CoCu111:b}}%
\phantomsubfloat{\label{fig:Vka_CoCu111:c}}%
\phantomsubfloat{\label{fig:Vka_CoCu111:d}}%
\phantomsubfloat{\label{fig:Vka_CoCu111:e}}% 
\phantomsubfloat{\label{fig:Vka_CoCu111:f}}%
\phantomsubfloat{\label{fig:Vka_CoCu111:g}}%
\phantomsubfloat{\label{fig:Vka_CoCu111:h}}%
\vspace{-2\baselineskip}% Remove extra line inserted by subfloat
\caption{(a) Density of states and (b) band structure of Cu(111). The shadowed regions correspond to the bulk bands projected into the surface. Two surface states are indicated, $SS_1$ (green) and $SS_2$ (red). 
(c)-(h) $|V_{n\mathbf{k}a}|^2$ for three orbitals of Co on Cu(111) on hollow position at $1.69$ \r{A}. The colors and line widths show the correspondence between the bands in (a) and the $|V_{n\mathbf{k}a}|^2$ in (f)-(h). The contour plots (c)-(e) show the $|V_{n\mathbf{k}a}|^2$ corresponding to the $SS_2$ surface state in the first Brillouin zone for the different orbitals. The $\mathbf{k}$-path K$\Gamma$MK shown in (b) and in (f)-(h)  is indicated in the contour plots.
\label{fig:Vka_CoCu111}}
\end{figure*}

Our approach to calculate  the Anderson self-energies allows us to clearly identify the contribution of the surface states that appear in the Cu(111) surface. 
In \cref{fig:Vka_CoCu111}  we show the surface band structure and the  square modulus of the atom-band couplings $|V_{n\mathbf{k}a}|^2$ computed with \cref{eq:vkas1} for the orbitals $d_{xz}$, $d_{z^2}$ and $s$.
The light shadowed regions correspond to the bulk Cu bands, projected into the (111) surface. 
They were obtained from a bulk calculation of the system, using as  supercell the same slab  as for the surface calculation, without the extra vacuum  between  slabs repetitions. 
For the bulk calculation we used the same $\mathbf{k}$-point  grid used in the slab calculation in the surface plane and a single $\mathbf{k}$-point in the normal [111] direction.

It is clear from the total density of states (DOS, \cref{fig:Vka_CoCu111:a}) that most of the states lie between $-5$ eV and $-1$ eV, corresponding to the $d$-bands of the bulk.
In \cref{fig:Vka_CoCu111:b}, we clearly identify the Shockley surface states (SS), $SS_1$ and $SS_2$. 
The $d$ orbitals  present a coupling with the surface states that is comparable or lower than the one  corresponding to the bulk bands (\cref{fig:Vka_CoCu111:c,fig:Vka_CoCu111:d,fig:Vka_CoCu111:f,fig:Vka_CoCu111:g}). 
The minor coupling of the $d$ orbitals with the SS has been used to  support  the hypothesis based on experimental data of a small influence of the surface states in the Kondo structures \cite{Wahl2004,Knorr2002}.
However, for the $s$ orbital (\cref{fig:Vka_CoCu111:e,fig:Vka_CoCu111:h}), the couplings with the Shockley surface states in the $sp$ band gap are the most important,  about ten times larger than the corresponding  couplings of the $d$ orbitals (note the change in scale). 
As we will discuss later on, this significant coupling of the surface states with the Co $4s$ orbital can explain the ZBA observed in the measured conductance spectra \cite{Knorr2002,Limot2005}.

The  $|V_{n\mathbf{k}s}|^2$ coupling of the  $s$ orbital with the surface states is strongly localized around the  $\Gamma$ point (\cref{fig:Vka_CoCu111:e,fig:Vka_CoCu111:h}).
Then, the  influence of these states in the hybridization function  will be mainly  seen at the energy of the SS bands close to that point of the reciprocal space (${\approx}{-0.5}$ eV).
The same is true for the $A1$ ($d_{z^2}$) orbital, which also  presents a localized coupling with the surface states around the $\Gamma$ point (\cref{fig:Vka_CoCu111:g}). 
For the $E1$ orbitals, there is a localized coupling with the Shockley states at the  K  points, where the corresponding bands  are at ${\approx}4.6$ eV (we only show  $d_{xz}$ in  \cref{fig:Vka_CoCu111:c,fig:Vka_CoCu111:f}). 
Then, we expect to observe an important influence of  the surface states in the $E1$ hybridization function at this energy.

\begin{figure}[ht]
\centering
\includegraphics[width=0.95 \linewidth,keepaspectratio]{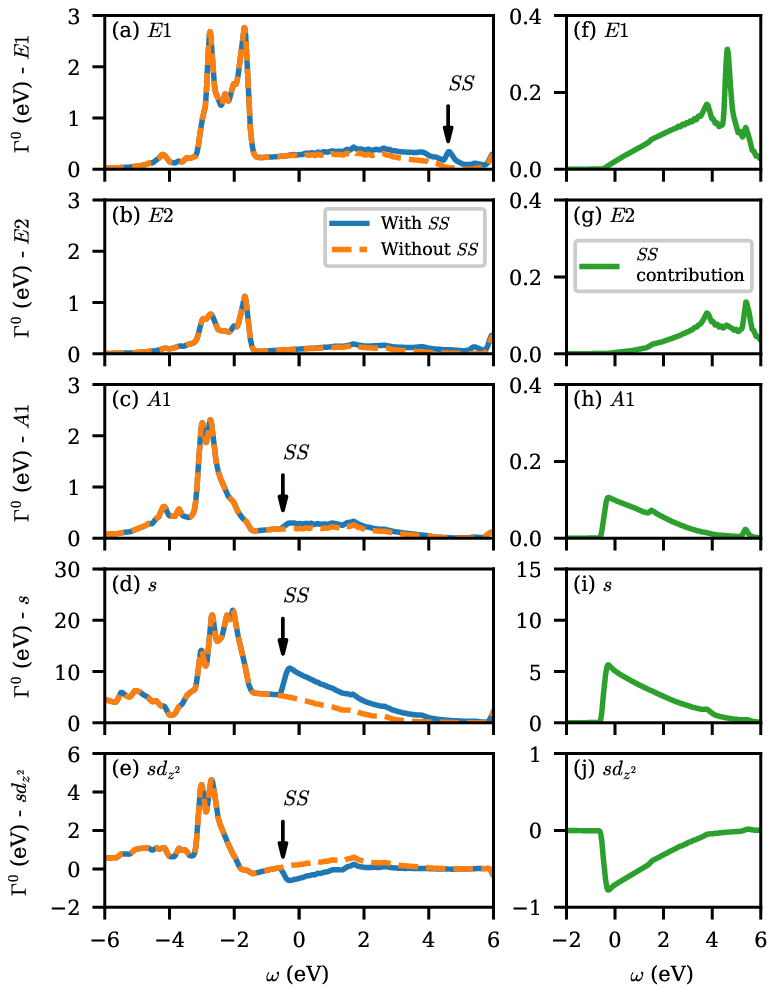}%
% Keep above the caption to avoid messing up counters
\phantomsubfloat{\label{fig:GammaCu_all:a}}%
\phantomsubfloat{\label{fig:GammaCu_all:b}}%
\phantomsubfloat{\label{fig:GammaCu_all:c}}%
\phantomsubfloat{\label{fig:GammaCu_all:d}}%
\phantomsubfloat{\label{fig:GammaCu_all:e}}%
\phantomsubfloat{\label{fig:GammaCu_all:f}}%
\phantomsubfloat{\label{fig:GammaCu_all:g}}%
\phantomsubfloat{\label{fig:GammaCu_all:h}}%
\phantomsubfloat{\label{fig:GammaCu_all:i}}%
\phantomsubfloat{\label{fig:GammaCu_all:j}}%
\vspace{-2\baselineskip}% Remove extra line inserted by subfloat
\caption{(a)-(e) Anderson widths $\Gamma^0$ for Co on Cu(111)   at the calculated adsorption height, including and neglecting the surface states $SS_1$ and $SS_2$ shown in \cref{fig:Vka_CoCu111}. 
The arrows indicate the energy position of the surface states bands in (a) the $K$  point and (c)-(e) the $\Gamma$ point.
(f)-(j) show the  contribution of the surface states to the  total Anderson widths. 
\label{fig:GammaCu_all}}
\end{figure}

In \cref{fig:GammaCu_all} we show  the Anderson widths 
$\Gamma^0 $ for each  orbital. 
The calculation  is performed by using \cref{eq:Gammadefinition,eq:vkas1}, where we identify the contribution of each band to the total hybridization functions. 
This approach  allows us to perform the theoretical exercise of neglecting the coupling of the surface states in the calculation of the Anderson self-energies. 
Then, in \cref{fig:GammaCu_all} we also present the results obtained  for Co on Cu(111) without including the  coupling with the Shockley states $SS_{1}$ and $SS_{2}$ in  \cref{eq:Gammadefinition}, in order to obtain an hypothetical Anderson width without the surface states contribution. 
In addition, we present the contribution corresponding to the surface states to the total widths.
Due to the symmetry of the system, only the $d_{z^2}$ orbital couples with the $s$ orbital  through the substrate band. 
Then, the only off-diagonal width that is non-null is 
$\Gamma^0_{sd_{z^2}}=\Gamma^0_{d_{z^2}s}$, shown in \cref{fig:GammaCu_all:e}.

The Anderson widths present a large weight between $\omega=-3$ eV and $\omega=-1$ eV, in the energy region where the bulk Cu $d$ bands are located (see \cref{fig:Vka_CoCu111:a}). 
Considering the position of the Shockley surface states in the Cu(111) band structure  and their coupling strength with the Co orbitals we can identify their influence on the atom-band hybridization functions $\Gamma^0$.  
In \cref{fig:GammaCu_all:a}, the $\Gamma_{E1}^0$ of Co on Cu(111) presents a peak at $\omega=4.6$ eV,  associated  to the coupling of the Shockley states  $SS_{1}$ and $SS_{2}$  with the Co $E1$  orbitals at the  $K$  points (\cref{fig:Vka_CoCu111:c,fig:Vka_CoCu111:f}). 
The same Shockley states are responsible for the structure  that appears in $\Gamma_{s}^0$  at $\omega=-0.5$ eV (\cref{fig:GammaCu_all:d,fig:GammaCu_all:i}), the energy position close to the $\Gamma$ point where  the surface states reach their maximum  coupling  with the Co $4s$ orbital  (\cref{fig:Vka_CoCu111:e,fig:Vka_CoCu111:h}). 
The couplings $\Gamma_{A1}^0$ and $\Gamma_{sd_{z^2}}^0$ are also  modified in the same energy region, while the influence on the remaining $E2$ orbitals close to the Fermi energy is negligible.

The ratio of the  contributions to the total hybridization function at the Fermi level of the surface ($\Gamma^{0(S)}$) and bulk states ($\Gamma^{0(B)}$),  $\Gamma^{0(S)}/\Gamma^{0(B)}$, corresponds to $0.08$ for the $E1$ orbitals, $0.03$ for the $E2$, $0.50$ for the $A1$ and $1.11$ for the $4s$ orbital. 
The values for the $d$ orbitals are roughly one order of magnitude larger than those  obtained in Ref. \cite{Barral2004}, where a   parametrized tight-binding Hamiltonian was used for their calculation. 
Our value for the ratio $\Gamma^{0(S)}/\Gamma^{0(B)}$ for the $A1$ orbital is instead in line  with an estimation based on modeling the quantum mirage  effect, which found  a lower limit of $0.1$ \cite{Aligia2005} and with a further refinement of the estimation leading to a value of  $1$   \cite{Moro-Lagares2018}, and agrees with the value used in Ref. \cite{Fernandez2021} to model the Co on Cu(111) system.

\subsection{Energy levels \label{sec:EnergyLevels}} 

The calculation of the energy levels is also performed  by using the bond-pair model \cite{Bolcatto1998}. The many-body Hamiltonian that describes the atom-surface interacting system is calculated in a mean-field approximation and introducing a second order expansion in the atomic overlap, in order to obtain the total energies of each configuration involved, which in turn define the energy levels active in the transitions  through the differences between them.

The asymptotic levels with respect to the vacuum were obtained by taking into account   experimental data  of the  excited neutral ($3d^8 4s^1$) and ionic ($3d^7 4s^1$) configurations energies \cite{Radzig1985}, so that the asymptotic energy levels $\epsilon_{d_i}$ with respect to vacuum are  given by $E(3d^7 4s^1)-E(3d^8 4s^1)=-7.76$ eV.
The image potential contribution, $1/4(z-z_p)$, in atomic units (a.u.) and at a normal distance to the surface  $z$  was taken into account by considering a matching distance of the long and  short range interactions of $z_c=8$ a.u. and the image plane at $z_p=2$ a.u.    \cite{Smith1989,Bolcatto1998}. 
The energy levels are then referred to the surface Fermi level,  considering a work function value  of  $4.94$ eV   \cite{Gartland1972}. 
The obtained values at the Co adsorption distance  were $\epsilon_{E1}=-2.6$ eV,  $\epsilon_{E2}=\epsilon_{A1}=-2.5$ eV and $\epsilon_{s}= 2.5$ eV.

\begin{figure}[h!]
\centering
\includegraphics[width=0.95\linewidth,keepaspectratio]{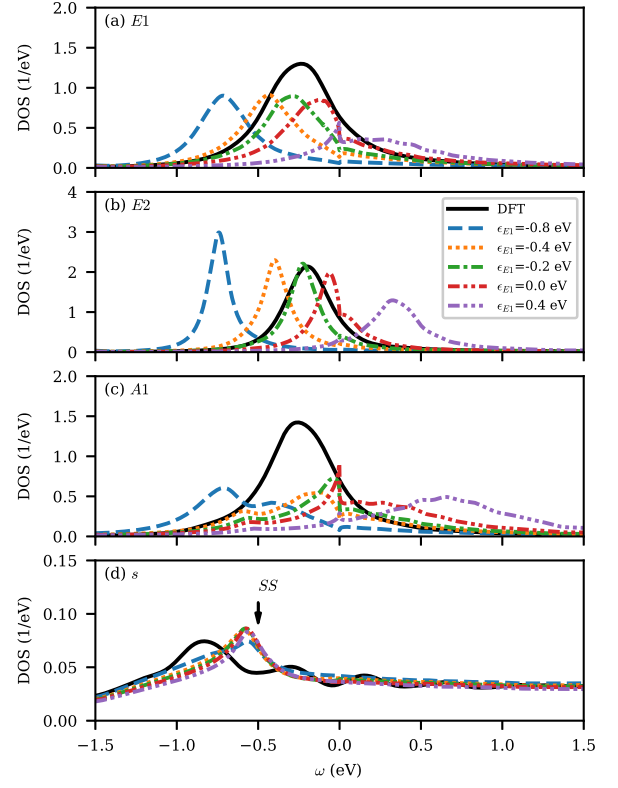}%
% Keep above the caption to avoid messing up counters
\phantomsubfloat{\label{fig:NoCorrelated_vs_DFT_varios:a}}%
\phantomsubfloat{\label{fig:NoCorrelated_vs_DFT_varios:b}}%
\phantomsubfloat{\label{fig:NoCorrelated_vs_DFT_varios:c}}%
\phantomsubfloat{\label{fig:NoCorrelated_vs_DFT_varios:d}}%
\phantomsubfloat{\label{fig:NoCorrelated_vs_DFT_varios:e}}%
\phantomsubfloat{\label{fig:NoCorrelated_vs_DFT_varios:f}}%
\phantomsubfloat{\label{fig:NoCorrelated_vs_DFT_varios:g}}%
\phantomsubfloat{\label{fig:NoCorrelated_vs_DFT_varios:h}}%
\phantomsubfloat{\label{fig:NoCorrelated_vs_DFT_varios:i}}%
\phantomsubfloat{\label{fig:NoCorrelated_vs_DFT_varios:j}}%
\vspace{-2\baselineskip}% Remove extra line inserted by subfloat
\caption{Spectral densities for each orbital group calculated with the NC model (discontinuous lines),  compared with DFT partial DOS (full lines). We performed calculations for different energy level positions,  assuming the  splittings obtained with the bond-pair calculation. The arrow  indicates the position of the Shockley states $SS$ at the $\Gamma$ point, which introduces the resonance-like peak in the $s$ orbital spectral density.
\label{fig:NoCorrelated_vs_DFT_varios}}
\end{figure}

The obtained energy levels of the $d$ orbitals lie  considerably below the Fermi level. As an alternative method to estimate the energy levels,  we use  a simplified   model (NC) which consists in calculating the spectral densities by disregarding multiorbital correlation. 
Then, we consider the orbitals as being independent one from each other, although we keep the normalization condition for the occupation probabilities. 
For the calculation of the Green functions of the $d$ orbitals under our NC approximation, we keep only the terms with $q'=q$ and $p'=p$ in  \crefrange{eq:eqdiagwiths4}{eq:eqdiagwiths5}. 
On the other hand, for the $s$ orbital we use   directly  the independent particle Green function, \cref{eq:Gss0}. 
In both cases, we obtain the spectral densities from the imaginary part of the corresponding Green functions by using \cref{eq:SpectralDensity}.
We compared the spectral densities obtained  by using the NC  model and several energy level shifts, maintaining the level splitting of the bond-pair calculation,  with the DFT partial DOS (PDOS). 
The results are presented in  \cref{fig:NoCorrelated_vs_DFT_varios}.

\Cref{fig:NoCorrelated_vs_DFT_varios} shows a good agreement between the PDOS calculated with DFT and the spectral densities of the NC model   when we use    the energy levels positions corresponding to $\epsilon_{E1} \approx  {-0.2}$ eV. 
Then, the values of $\epsilon_{E1}$ for which the NC results approximately match the PDOS are shifted ${\approx}  2.4$ eV  with respect to the values given by the  bond-pair model. 
This kind of uncertainty in the energy levels is present in many approaches used to compute them  \cite{Mozara2018,Jacob2010,Baruselli2015,Wehling2010}, and  rigid shifts on  the  energy levels or chemical potential are usually introduced to improve the description of experimental results. 
Therefore, we  will present calculations for different energy levels, considering  them as adjustable parameters of our model.
We use the criterion of maintaining the splitting between levels obtained with the bond-pair model. In this way,  we consider the energy splittings $\epsilon_{E2}-\epsilon_{E1}= \epsilon_{A1}-\epsilon_{E1}=0.1$ eV and $\epsilon_{s}-\epsilon_{E1}=5.0$ eV, 
and we use as reference the position of the $\epsilon_{E1}$ energy level.

We should notice that although our NC model   neglects   correlation between  multiple configurations, it includes correlation in each  orbital, which introduces the  structures observed in the spectral densities at the Fermi level  (\crefrange{fig:NoCorrelated_vs_DFT_varios:a}{fig:NoCorrelated_vs_DFT_varios:c}).

In \cref{fig:NoCorrelated_vs_DFT_varios:d}, the $\epsilon_s$ energy level position varies between $4.2$ eV and $5.4$ eV.
In contrast with the $d$ orbitals results, the  spectral densities corresponding to the $s$ orbital, presented in \cref{fig:NoCorrelated_vs_DFT_varios:d}, remain largely unaffected by the energy level  shifts.
In addition, they are  mainly structureless, with the exception of a peak at $\omega \approx -0.5$ eV observed in the NC calculation.
The peak appears to be shifted to lower energies in the DFT PDOS. 
This peaked structure, introduced by the Anderson self-energy in the independent particle Green function of \cref{eq:Gss0}, is related to the Shockley states $SS_{1}$ and $SS_{2}$, and can be also noticed (in a smaller scale) in the spectral density corresponding to the $A1$ orbital,  \cref{fig:NoCorrelated_vs_DFT_varios:c}.

Notice that the DFT results presented in \cref{fig:NoCorrelated_vs_DFT_varios} use a supercell approach for the calculation, which would require a much larger cell size  to properly describe the surface states of Cu(111) in presence of the impurity \cite{Barral2004}. 
On the other hand, the self-energies that we calculated are computed from the data of the Cu surface  without the impurity, so that they  include the surface states given by the DFT calculation of a $1 \times 1$ cell of a clean Cu(111) slab.

\section{Correlated calculations \label{sec:conductance}}

\subsection{Occurrence probabilities of each configuration and orbital occupations}

\begin{figure}[ht]
\centering
\includegraphics[width=0.99 \linewidth,keepaspectratio]{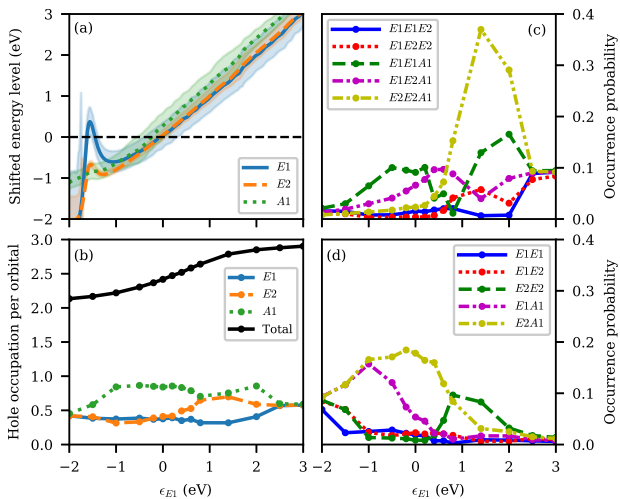}%
% Keep above the caption to avoid messing up counters
\phantomsubfloat{\label{fig:Occupation_levels:a}}% 
\phantomsubfloat{\label{fig:Occupation_levels:b}}%
\phantomsubfloat{\label{fig:Occupation_levels:c}}%
\phantomsubfloat{\label{fig:Occupation_levels:d}}%
\vspace{-2\baselineskip}% Remove extra line inserted by subfloat
\caption{(a) Energy levels shifted by the real part of the Anderson self-energy, $\epsilon_{D}+\Lambda^0_D(\epsilon_{D})$, and broadened by its imaginary part, $\Gamma^0_D(\epsilon_{D})$, for each group $D=E1$, $E2$, $A1$. 
(b) Hole occupation per orbital for each group and total hole occupation in the $d$-shell. 
(c) Occurrence probabilities for each configuration with three holes, without considering the sum over equivalent  configurations. (d) Same as (c) for the two-holes configurations.
The results in  (b)-(d) take into account the summation over the spin projection index.
\label{fig:Occupation_levels}}
\end{figure}

In \cref{fig:Occupation_levels:a}  we present the one-particle energy levels shifted by the real part of the non-interacting self-energies, $\epsilon_{D}+\Lambda^0_D(\epsilon_{D})$ ($\Sigma_D^0=\Lambda^0_D+i\Gamma_D^0$, see \cref{eq:SelfEnergy}). The energy levels,  broadened by the imaginary part $\Gamma^0_D(\epsilon_{D})$, are presented for each group $D=E1$, $E2$, $A1$. 
We show the results as a function of the position of  $\epsilon_{E1}$, taken as reference and  being the relative energy positions of the other orbitals given by the
splitting obtained from the bond-pair model in \cref{sec:EnergyLevels}. 
The hole occupations and occurrence probabilities, which depend on the positions of the energy levels, are shown in the remaining panels of \cref{fig:Occupation_levels} as a function of the same reference value $\epsilon_{E1}$.

Our ionic Hamiltonian approach allows us to identify the occurrence probability of each atomic configuration of the considered space (see \cref{tab:PQfluctuations}). 
\Cref{fig:Occupation_levels:c} presents the occurrence probabilities for the configurations with three holes, obtained from  \cref{eq:occupation:p} and summed over the total spin projection index $M$. The same is shown in \cref{fig:Occupation_levels:d} for the two-holes configurations, obtained from \cref{eq:occupation:q}.
Note that, for example, the two-holes configuration $E1A1$ of \cref{fig:Occupation_levels:c} corresponds to the configuration  with holes in the $d_{z^2}$ orbital and either the $d_{xz}$ or the $d_{yz}$ orbital. That is, the occurrence probabilities of all the equivalent configurations are not summed in  \cref{fig:Occupation_levels:c,fig:Occupation_levels:d}  (see \cref{ap:configurations}).
The orbital occupations are obtained from the occurrence probabilities of each configuration by using \cref{eq:orbitaloccupation}. We present the hole occupation per orbital in \cref{fig:Occupation_levels:b}, together with the total hole occupation of the $d$-shell.

When $\epsilon_{E1} \lesssim -2$ eV all the orbital energies are well defined below the Fermi energy $E_F=0$ eV (\cref{fig:Occupation_levels:a}). Then, the occurrence probabilities of the three-holes configurations are negligible (\cref{fig:Occupation_levels:c}), and all the two-holes configurations are equally probable (\cref{fig:Occupation_levels:d}). 
As a result, the occupation of the five $d$ orbitals is the same (\cref{fig:Occupation_levels:b}) and the total hole occupation approaches to the value $2$.
The other limit situation, with all the energy levels well defined above $E_F$ occurs for   $\epsilon_{E1} \gtrsim 3$ eV. In this case, the three-holes configurations result the most probable, and the total hole occupation approaches to $3$.
For  $\epsilon_{E1}\gtrsim-1$ eV,  the shifted energy levels (\cref{fig:Occupation_levels:a}) increase quite linearly with the bare energy level $\epsilon_{E1}$. 

When the energy levels are close to the Fermi level (${-0.5 \text{ eV}} \lesssim {\epsilon_{E1}} \lesssim  {0.5 \text{ eV}}$), the three-holes configurations that have the $A1$ orbital occupied, namely $E1E1A1$, $E1E2A1$ and $E2E2A1$,  are the most probable  (\cref{fig:Occupation_levels:c}). 
The dominant two-holes configurations in this region are also those with the $A1$ orbital occupied, $E1A1$ and $E2A1$  (\cref{fig:Occupation_levels:d}).  
In a first approximation, we can relate this observation with the shifted energy levels presented in \cref{fig:Occupation_levels:a}. 
The $A1$ energy level remains above the other levels, and therefore is the most favorable orbital to be occupied by holes. 
As a result, the mean occupation of the $A1$ orbital including spin, approximately constant around $E_F$,  results of ${\approx} 0.85$ holes, that is ${\approx} 1.15$ electrons.
% and pushed towards a unit occupation with respect to the DFT result presented in \cref{tab:DFTOrbitalOccupation}.
We observe that the occupation of the $E1$ orbitals also presents a small variation in this region, which  can be related to the  relatively broad  energy  width  (\cref{fig:Occupation_levels:a}). 
On the other hand, the $E2$ level has a smaller Anderson width, giving place to a more marked dependence of the orbital hole occupation when the energy level is close to $E_F$. The hole occupation  tends to increase as the $E2$ level crosses $E_F$ (\cref{fig:Occupation_levels:b}). 
Around $E_F$, both $E1$ and $E2$ present a mean occupation of ${\approx} 0.4$ holes, that is ${\approx} 1.6$ electrons.
While the occupation of the $E1$ and $E2$ orbitals are similar to the DFT result presented in \cref{tab:DFTOrbitalOccupation} (${\approx} 1.55$ electrons), the occupation of the $A1$ orbital is driven closer to $1$ when compared with the DFT result of $1.54$ electrons.

Our ionic Hamiltonian proposal, where all the $d$ orbitals are considered as active, gives place to a multi-configurational state with many $S=3/2$ and $S=1$ configurations. 
For each three-holes configuration we have the possibility of fluctuations to three configurations with two holes (see \cref{tab:PQfluctuations:c}). 
From \cref{fig:Occupation_levels:c,fig:Occupation_levels:d} we can extract the occurrence probability of each three-holes configuration and the corresponding one for each two-holes configuration to which it can fluctuate, as a function of the energy level position. 
In principle, different probabilities suggest different correlation regimes associated to each fluctuation. If the $S=3/2$ configuration has a much larger probability, a Kondo regime is expected, while if the probabilities of both spin configurations, $S=3/2$ and $S=1$, are similar a mixed valence is occurring. Finally, the empty orbital regime takes place for a predominant probability of the $S=1$ configuration. 
On the other hand, if the three possible fluctuations for a given three-holes configuration correspond to the Kondo regime with very similar Kondo scales, a full screened Kondo effect is suggested. In our case, where different energy levels and hybridization widths are involved, either a partially screened or a two stage Kondo effect would be expected \cite{Nozieres1980,Posazhennikova2007}.   

It is important to note that in our proposal all the configurations become mixed, and the resulting charge redistribution that occurs can suppress the local moment of the Kondo-active space of configurations  \cite{Valli2020}. 
As it was discussed in a previous work \cite{Tacca2020}, the spectral densities calculated with our model show various peaks whose positions and widths are determined by the self-energies  defined by \cref{eq:eqdiagwiths4}. These structures, coming from the different orbitals that give place to virtual transitions, are  related to the electronic correlation in a multiorbital system \cite{Tacca2020}.

\subsection{Considerations for  conductance calculations}

\begin{figure}[ht]
\centering
\includegraphics[width=0.8 \linewidth,keepaspectratio]{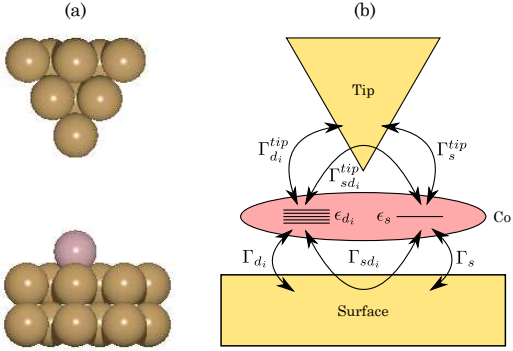}%
\caption{(a) Geometry  and (b) couplings involved in the surface-atom-tip system.
\label{fig:esquema_tip}}
\end{figure}

We  consider the conductance through a  Co adatom on Cu(111) when a tip is placed on top of it. 
With that aim we modeled a tip of Cu  with a three layers pyramid  on a $4\times4\times6$ slab of Cu(111). 
The calculation of the atom-tip Anderson widths $\Gamma^{0\text{-}tip}$ was done following the same procedure of  \cref{sec:HamiltonianParameters}, by  computing the dimeric couplings between the Co adatom and the atoms of the tip and afterwards  constructing  $\Gamma^{0\text{-}tip}$ by using the density matrix of the Cu tip obtained by DFT.
In \cref{fig:esquema_tip} we present schemes of the geometry of the surface-atom-tip system  and the hybridization functions involved.

\begin{figure}[ht] 
\centering
\includegraphics[width=0.85 \linewidth,keepaspectratio]{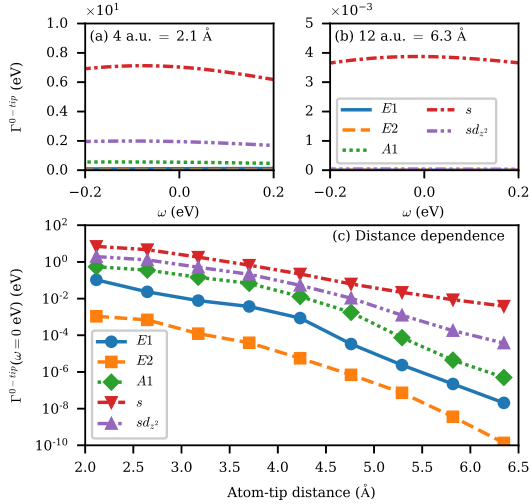}% 
% Keep above the caption to avoid messing up counters
\phantomsubfloat{\label{fig:GammaTip:a}}% 
\phantomsubfloat{\label{fig:GammaTip:b}}%
\vspace{-2\baselineskip}% Remove extra line inserted by subfloat
\caption{Atom-tip coupling  $\Gamma^{0\text{-}tip}$ for  Co-tip distances of  (a) $2.1$ \r{A} and (b) $6.3$ \r{A}, for each orbital. (c) shows the dependence of $\Gamma^{0\text{-}tip}$ at the Fermi level with the atom-tip  distance. 
\label{fig:GammaTip}} 
\end{figure}

The hybridization functions of the Co adatom with the tip are shown in \cref{fig:GammaTip}. We present the dependence of $\Gamma^{0\text{-}tip}$ evaluated at the Fermi level with the atom-tip distance, as well as $\Gamma^{0\text{-}tip}(\omega)$ when the tip is near ($2.1$ \r{A}) and far ($6.3$ \r{A}) from the Co adatom.

The hybridization strength follows symmetry considerations: the extended $s$ orbital has  the largest coupling, and the  $A1$ orbital is favored because the tip is on top of the adatom. 
The $E1$ orbitals have a considerably lower coupling, while the coupling of  the $E2$ orbitals, located in the surface plane, is negligible.  
At large atom-tip distances, the relative coupling of the $s$ orbital with respect to the remaining ones increases. 
Then, when the tip is far from the atom, 
the transmission through the $d$ orbitals (\cref{eq:Td}) is negligible due to their localization  \cite{Merino2004}, and the  transmission will occur mainly  through the $s$ orbital (\cref{eq:Ts}). 
In this case, if there is a zero-bias anomaly (ZBA)  in the conductance spectra introduced by the Co $4s$ conduction channel, this will be the dominant one.

We should mention that in our calculations we considered the Co adatom at its equilibrium position without the tip, that is, we used the hybridization functions presented in \cref{sec:SelfEnergies} for the atom-surface coupling and did not compute a relaxed geometry including the tip at different distances. 
This approximation is supported by the observation that  geometry relaxation effects can be neglected in this system \cite{Vitali2008,Baruselli2015,Barral2004}.

\subsection{Effect of the $3d_{z^2}$ orbital in the Co $4s$ spectral density and surface states influence \label{sec:LorentzianG}}

The conductance of the $s$ orbital, calculated from \cref{eq:Ts}, has two contributions: the independent particle term given by $G_{ss}^0$ (\cref{eq:Gss0}) and the term introduced by the interaction with the correlated $d$ orbitals, $G_{s}^{(c)}$ (\cref{eq:Gsc} \cite{Calvo2012}. Then, the $T_s$ contribution to the conductance is given by
\begin{equation}  \label{eq:Ts1}
T_{s}(\omega) = 
  2 \Gamma_{s}^{eff}(\omega) \text{Im}G_{ss}^{0}(\omega) + \Gamma_{s}^{eff}(\omega) \text{Im} G_{s}^{(c)}(\omega) .
\end{equation}

Clearly, any possible zero-bias anomaly in the conductance spectra of the $s$ orbital will be introduced by the second term of \cref{eq:Ts1}, given that the first term corresponds to an independent particle  calculation. 
The function $G_{s}^{(c)}$  is given by \cref{eq:Gsc,eq:sigma}, and depends on the off-diagonal Anderson widths $\Gamma_{sd_i}^0$.
Given that, among the off-diagonal hybridization functions between the $s$ and $d$ orbitals, only $\Gamma_{sd_{z^2}}$ is non-zero, we obtain
\begin{equation}  \label{eq:Ts2}
\text{Im} G_{s}^{(c)}(\omega) =  
  \text{Im}\left(\left(\sigma_{sd_{z^2}}(\omega)\right)^{2}G_{d_{z^2}}(\omega)\right) ,
\end{equation}
where 
\begin{equation} \label{eq:sigma2}
 \left(\sigma_{sd_{z^2}}(\omega)\right)^{2}=\left(\frac{\Sigma_{sd_{z^2}}^{0}(\omega)}{\omega-\epsilon_{s}-\Sigma_{s}^{0}(\omega)}\right)^{2}
 .
\end{equation}
In this way, $\sigma_{sd_{z^2}}$  gives place to an interference between the real and imaginary parts of the $G_{d_{z^2}}$ Green function, and can introduce structures in the conductance spectra of the $s$ level.

Let us analyze the effect of $\sigma_{sd_{z^2}}$ in $T_s$. With that aim, we  assume that 
$G_{d_{z^2}}$  is given by an hypothetical Green function $G_d$ \cite{Calvo2012},
\begin{equation} \label{eq:GdKondo}
 G_d(\omega)=
 \frac{Z}{\omega-i \Gamma_\mathrm{K}}=
 \frac{Z/\Gamma_{\mathrm{K}}}{\left(\omega/\Gamma_{\mathrm{K}}\right)^{2}+1} \left(
 \omega/\Gamma_{\mathrm{K}} 
 +i   \right),
\end{equation}
the imaginary part of which corresponds to a Lorentzian peak of width $\Gamma_\mathrm{K}$ centered at $\omega=0$ eV. Using \cref{{eq:GdKondo}} for  $G_d$, we evaluate the  structures that are introduced in $T_s$ (\cref{eq:Ts2})  through  
$\left( \sigma_{sd_{z^2}} \right)^2$, which is in turn calculated from the self-energies  using  \cref{eq:sigma2}. 
We can write the contribution of $\text{Im} G_{s}^{(c)}(\omega)$ in \cref{eq:Ts2} as a Fano-like function \cite{Fano1961,Calvo2012}. By replacing \cref{eq:GdKondo} in \cref{eq:Ts2} we obtain  
\begin{equation} \label{eq:Gsscc}
 \text{Im}G_{s}^{(c)}(\omega)
 =\frac{Z}{\Gamma_{\mathrm{K}}}\left(\text{Im}\sigma_{sd_{z^2}}(\omega)\right)^{2}\left(\frac{\left(q_F+\omega/\Gamma_{\mathrm{K}} \right)^{2}}{1+(\omega/\Gamma_{\mathrm{K}})^{2}}-1\right), 
\end{equation}
where  the Fano factor $q_F$ is defined as
\begin{equation} \label{eq:FanoFactor}
 q_F=\frac{\text{Re} \sigma_{sd_{z^2}}(\omega) }{\text{Im} \sigma_{sd_{z^2}}(\omega) }.
\end{equation}

\begin{figure}[h!]
\centering
\includegraphics[width=0.89\linewidth,keepaspectratio]{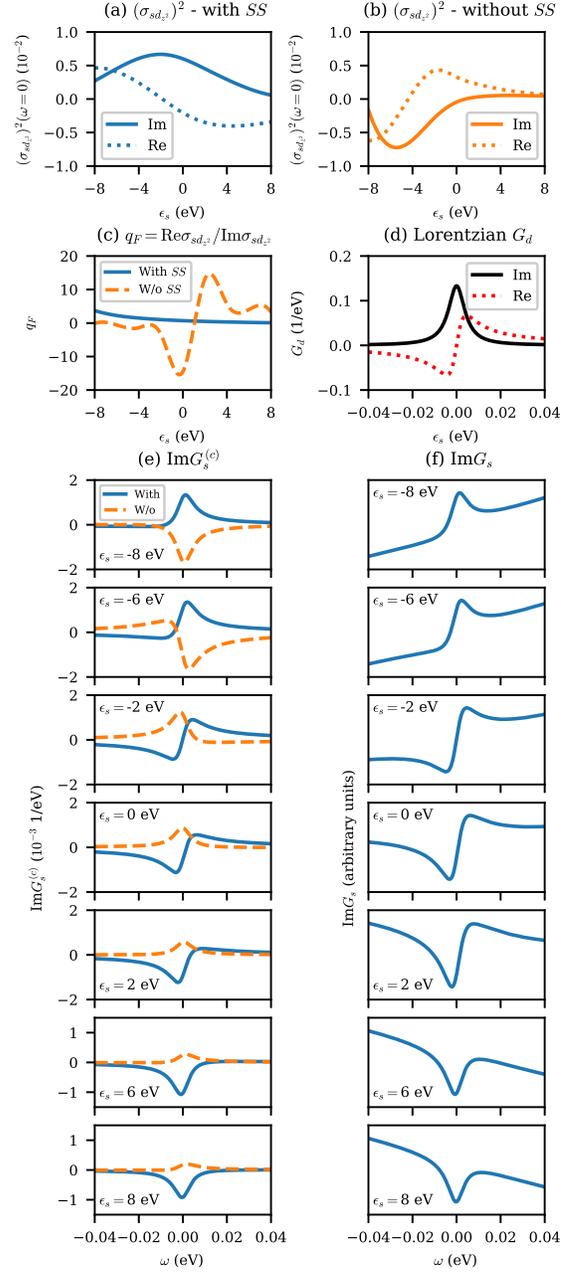}% 
% Keep above the caption to avoid messing up counters
\phantomsubfloat{\label{fig:SpectralDensityGs_Kondo_juntos:a}}%
\phantomsubfloat{\label{fig:SpectralDensityGs_Kondo_juntos:b}}%
\phantomsubfloat{\label{fig:SpectralDensityGs_Kondo_juntos:c}}%
\phantomsubfloat{\label{fig:SpectralDensityGs_Kondo_juntos:d}}%
\phantomsubfloat{\label{fig:SpectralDensityGs_Kondo_juntos:e}}%
\phantomsubfloat{\label{fig:SpectralDensityGs_Kondo_juntos:f}}%
\vspace{-2\baselineskip}% Remove extra line inserted by subfloat
\caption{%
(a) Quantity $\left( \sigma_{sd_{z^2}} \right)^2$ calculated using \cref{eq:sigma2} from the obtained self-energies and evaluated at $\omega=0$ eV, as a function of  $\epsilon_s$. 
(b) The same quantity than in (a) but computed from the self-energies without including the surface states.
(c) Fano factor $q_F$, calculated using \cref{eq:FanoFactor} and as a function of $\epsilon_s$, including or neglecting the surface states contribution.
(d) Hypothetical $G_d$ given by \cref{eq:GdKondo},  with Lorentzian imaginary part.
The panels of (e) show the contribution to the $s$ spectral density given by the interaction with the $d_{z^2}$ orbital via the surface band, $\mathrm{Im}G_s^{(c)}$, considering the theoretical $G_d$ of panel (d) and the $\left( \sigma_{sd_{z^2}} \right)^2$ quantity, for several values of   $\epsilon_s$ and including (full lines) or neglecting (dashed lines) the surface states contribution.
(f) corresponds to the total $\mathrm{Im}G_s$, which, in addition to the contribution of (e),  includes  the independent electron contribution, $\mathrm{Im} G_{ss}^0$. 
\label{fig:SpectralDensityGs_Kondo_juntos}}
\end{figure}

Notice that the Fano structure of \cref{eq:Gsscc} has the same width $\Gamma_{\text{K}}$  as the Lorentzian $G_d$ of \cref{eq:GdKondo}. 
Then, the resonance  width $\Gamma_{\text{K}}$, associated with the Kondo temperature, is determined by the original Kondo resonance that emerges from the correlated $d$ orbitals, which hybridize mainly with the bulk states. 
The effect of the surface states is to modify the line shape by means of the interaction with the $s$ orbital.
We present the results of this analysis 
in \cref{fig:SpectralDensityGs_Kondo_juntos}.
Given that it is observed a negligible dependence of $\sigma_{sd_{z^2}}(\omega)$ with $\omega$ in the region near the Fermi level where the interference structures are relevant  
($|\omega| \lesssim  0.04$ eV),  we consider its value at $\omega=0$ eV to calculate $q_F$.

\Cref{fig:SpectralDensityGs_Kondo_juntos:a} shows the parameter 
$\left(\sigma_{sd_{z^2}}\right)^2$ evaluated at $\omega=0$ eV and as a function of the $s$ orbital energy level, $\epsilon_s$. \Cref{fig:SpectralDensityGs_Kondo_juntos:b} presents the same quantity but without including the surface states in the calculation of the self-energies (see \cref{fig:GammaCu_all}).
In \cref{fig:SpectralDensityGs_Kondo_juntos:c} the $q_F$ factor (\cref{eq:FanoFactor}) is shown also as a function of $\epsilon_s$, including and neglecting the surface states contribution.
In \cref{fig:SpectralDensityGs_Kondo_juntos:d}, we show the theoretical $G_d$ given by \cref{eq:GdKondo}. We used $\Gamma_\mathrm{K}=4.5 \times 10^{-3}$ eV, which corresponds to the Kondo temperature  $T_{\mathrm{K}} \approx 54$ K estimated from measurements of Co on Cu(111) \cite{Knorr2002} and we chose $Z=6\times 10^{-4}$, based on the scale of the Kondo structures obtained in \cref{sec:conductancespectra}. We notice that this $Z$ value is lower than the one expected by the estimation $Z=\Gamma_\mathrm{K} / \Gamma_{A1}(E_F)\approx 1.5 \times 10^{-2}$. This underestimation of the Kondo structure is a known artifact of the second order approximation of the EOM \cite{Monreal2005}.
The real and imaginary parts of $\left(\sigma_{sd_{z^2}}\right)^2$ vary with $\epsilon_s$ and, together with $G_d$, produce different interference structures, introduced by $\mathrm{Im}G_s^{(c)}$  in the spectral density of the Co $4s$ orbital. 
The  $\mathrm{Im}G_s^{(c)}$ contribution to the spectral density is shown in \cref{fig:SpectralDensityGs_Kondo_juntos:e}, for different values of $\epsilon_s$ and including or neglecting the surface states contribution. 
The shape of the interference structures respond to the $q_F$ factor  shown in \cref{fig:SpectralDensityGs_Kondo_juntos:c}.
In \cref{fig:SpectralDensityGs_Kondo_juntos:f}, the contribution of the independent electron Green function 
$\mathrm{Im}G_{ss}^0$ is added to  $\mathrm{Im}G_s^{(c)}$, in order to obtain the total $\mathrm{Im}G_s$, which defines the spectral density. 

We analyze first the results obtained when  we include the surface states in our calculation of the self-energies. 
We observe that $\text{Re}((\sigma_{sd_{z^2}})^2) \approx \text{Im}((\sigma_{sd_{z^2}})^2)$ when $\epsilon_s\approx -6$ eV  (\cref{fig:SpectralDensityGs_Kondo_juntos:a}), so that the structure has similar contributions from the real and imaginary parts of $G_d$ at this value of $\epsilon_s$. 
This situation corresponds to a Fano factor $q_F \approx 2$ (\cref{fig:SpectralDensityGs_Kondo_juntos:c}), giving rise to a slightly asymmetric line shape (\cref{fig:SpectralDensityGs_Kondo_juntos:e}). 
The relation changes when $\epsilon_s$ is shifted to higher energies. In particular, at $\epsilon_s=-2$ eV,   $\text{Re}((\sigma_{sd_{z^2}})^2)$ becomes negative, so that the peak in $\text{Im}G_d$ is transformed into a dip. 
The dip is better defined when $\epsilon_s$ is further shifted to higher energies, given that $\text{Re}((\sigma_{sd_{z^2}})^2)$ approaches zero. In the same way,  $q_F$ tends to zero when 
$\epsilon_s$ is increased, in correspondence with the dip-like shape.
By comparing  \cref{fig:SpectralDensityGs_Kondo_juntos:e} with \cref{fig:SpectralDensityGs_Kondo_juntos:f},  we can observe that the effect of the independent electron Green function $G_{ss}^0$  is to add an approximately linear background to the total spectral density of the $s$ orbital. The slope of this contribution is positive for $\epsilon_s \lesssim -2$ eV, and negative for $\epsilon_s  \gtrsim 2$ eV.

We can find the value of $\epsilon_s$ that gives the Fano factor $q_F$ extracted from experimental measurements of the conductance of Co on Cu(111), that is, $q_F=0.18\pm0.03$ \cite{Knorr2002}. By using  \cref{eq:FanoFactor}, we find that $\epsilon_s=6.2$ eV leads to $q_F=0.18$. 
The $\epsilon_s$ level position obtained is close to our estimation given by the bond-pair calculation of the energy level including the shift suggested by DFT results, which positioned the $\epsilon_s$ level above the Fermi level at $\epsilon_s \approx 5$ eV (see \cref{sec:EnergyLevels}). 
This position of the energy level gives an occupation of the  $s$ orbital  (\cref{eq:occs}) of $0.18$ electrons, which is also consistent with our DFT results presented in \cref{tab:DFTOrbitalOccupation} ($s$ orbital occupation of $0.25$). 

We evaluate now how our results change when we neglect  the surface states contribution.
It has been found that a proper description of the  surface states can be relevant in the calculation of the conductance spectra of adatoms in metallic (111) surfaces \cite{Moro-Lagares2018,Merino2004}.
In \cref{fig:SpectralDensityGs_Kondo_juntos:c} we   observe that the change in the Anderson widths induced by the exclusion of the surface state  couplings strongly modifies the Fano factor $q_F$  that determines the interference line shape of $\text{Im}G_s^{(c)}$ shown in  \cref{fig:SpectralDensityGs_Kondo_juntos:e}. 
When the $s$ energy level is located at $2 \text{ eV}  \lesssim \epsilon_{s} \lesssim 8 \text{ eV}$, the dip observed when the  SS  are included  changes to a peak when their contribution is neglected (\cref{fig:SpectralDensityGs_Kondo_juntos:e}), corresponding to a Fano factor $q_F>1$ (\cref{fig:SpectralDensityGs_Kondo_juntos:c}). 
Between  $-2 \text{ eV}  \lesssim \epsilon_{s} \lesssim 1 \text{ eV}$, the Fano factor is  $q_F<-1$  and a peak structure is also observed in the case of neglecting the couplings with the Shockley states (\cref{fig:SpectralDensityGs_Kondo_juntos:e}).
Only for a large negative value of $\epsilon_{s} =-8$ eV we obtain a dip-like structure in the case of disregarding the presence of the surface states. This position of the energy level $\epsilon_{s} $, well below the Fermi level, can not be justified. 
In summary, if we do not consider the coupling of the Co adatom with the surface states of Cu(111), the dip-like structure  obtained at reasonable values of $\epsilon_{s} $ is transformed into a peak-like structure. 
This result shows that the localized surface states play a key role in the conductance  spectra through the strong hybridization with the Co $4s$ orbital. Then, a correct description of the surface states is especially important in this system. 

\subsection{Conductance spectra and dependence with the atom-tip distance  \label{sec:conductancespectra}}

\begin{figure}[ht]
\centering 
\includegraphics[width=0.95\linewidth,keepaspectratio]{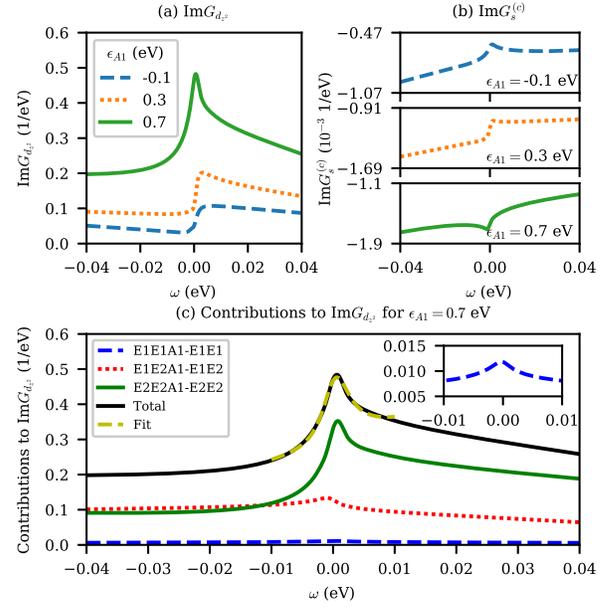}%  
% Keep above the caption to avoid messing up counters
\phantomsubfloat{\label{fig:GdGs_zoom_corr:a}}%
\phantomsubfloat{\label{fig:GdGs_zoom_corr:b}}% 
\phantomsubfloat{\label{fig:GdGs_zoom_corr:c}}%
\vspace{-2\baselineskip}% Remove extra line inserted by subfloat
\caption{(a) Correlated Im$G_{d_{z^2}}$, for different energy level positions.(b) Resulting $\text{Im}G_s^{(c)}$ given by \cref{eq:Gsscc}. 
(c) Contributions to Im$G_{d_{z^2}}$ at $\epsilon_{A1}=0.7$ eV of the possible transitions involving $A1$ as active orbital.   A fit of the total Im$G_{d_{z^2}}$ using a Fano function and is shown. The inset shows a zoom around $\omega=0$ eV to appreciate the $E1E1A1$-$E1E1$ contribution. 
\label{fig:GdGs_zoom_corr}}
\end{figure}

We continue now with the  Green functions given by our correlated calculation. The calculations were made with the energy splitting calculated by using  the bond-pair model, presented in \cref{sec:EnergyLevels}, and for different energy shifts. 
We considered a temperature $T=4.2$ K.  
In \cref{fig:GdGs_zoom_corr} we present the resulting 
$\text{Im}G_{d_{z^2}}$ and $\text{Im}G_s^{(c)}$, for energy level positions between $\epsilon_{E1}=-0.2$ eV and $\epsilon_{E1}=0.6$ eV, that is, between $\epsilon_{A1}=-0.1$ eV and $\epsilon_{A1}=0.7$ eV.  
Recall that the value  $\epsilon_{E1}=-0.2$ eV was suggested by the comparison between our NC model and DFT results in \cref{sec:EnergyLevels}. 

We  see in  \cref{fig:GdGs_zoom_corr:a} that  we obtain  a better defined structure when we shift the energy level positions according   to $\epsilon_{E1}=0.6$ eV, which corresponds to $\epsilon_{A1}=0.7$ eV.  
Panel  \subref{fig:GdGs_zoom_corr:b}  shows the  different structures in   $\text{Im}G_s^{(c)}$ for the energy level positions considered. 
The structures vary with the energy levels, and change from a peak-like to a dip-like shape.
Taking into account the experimental line shape  \cite{Limot2005,Knorr2002}, our calculations at  the energy level positions corresponding to  $\epsilon_{A1}=0.7$ eV reproduce qualitatively the experimental results. 
The better defined Kondo structure obtained introducing  this shift to the energy levels resembles a Lorentzian peak, similar to the $G_d$   used in  \cref{fig:SpectralDensityGs_Kondo_juntos}. 
The shift of the energy levels to  $\epsilon_{A1}=0.7$ eV also leaves the $4s$ orbital energy at $5.6$ eV, closer to the value suggested by the experimental $q_F$ in \cref{sec:LorentzianG}. 

In  \cref{fig:GdGs_zoom_corr:c} we present the contributions of the possible fluctuations to Im$G_{d_{z^2}}$, considering  
$\epsilon_{A1}=0.7$ eV. 
The occurrence probabilities of the corresponding configurations involved in the three fluctuations (see \cref{fig:Occupation_levels:c,fig:Occupation_levels:d}) suggest a Kondo regime for $E1E1A1$-$E1E1$ and $E1E2A1$-$E1E2$, and for $E2E2A1$-$E2E2$ a Kondo regime towards a mixed valence one.
The structures shown in \cref{fig:GdGs_zoom_corr:c} indicate that the Kondo resonance is mainly defined by the fluctuation between the configurations $E2E2A1$ and $E2E2$. 
This contribution introduces an asymmetric structure, which is in turn translated into  asymmetric structures in Im$G_{s}^{c}$ in \cref{fig:GdGs_zoom_corr:b}.
We should note that our description of the Kondo resonance is certainly limited by our approximation based on the EOM method closed up to a second order in the atom-band coupling. 
In \cref{fig:GdGs_zoom_corr:c}, we fit a Fano function \cite{Fano1961} to the result obtained at $\epsilon_{A1}=0.7$ eV,  
\begin{equation} \label{eq:Fano}
 f_{FA}(\epsilon)=\frac{1}{1+q_F^2} \left( \frac{\left(q_F+\epsilon \right)^{2}}{1+\epsilon^{2}} - 1 \right) ,
\end{equation}
considering a linear offset, an energy window of $\pm0.01$ eV and with $\epsilon= (\omega-\omega_0)/(k_B \Gamma_{FA}) $. 
The fitting parameter $\omega_0$ corresponds to the center of the resonance, while $k_B$ is the Boltzmann constant. 
The resulting  half width at half maximum $\Gamma_{FA}=31.4 \pm 0.6 $ K provides  an estimation of the Kondo temperature \cite{Gruber2018,Daroca2018}, which is in  reasonable agreement with experimental  data \cite{Knorr2002}.

\begin{figure}[ht]
\centering 
\includegraphics[width=0.8\linewidth,keepaspectratio]{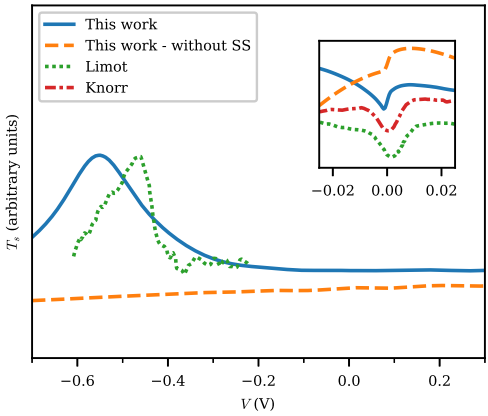}%
% Keep above the caption to avoid messing up counters
% \phantomsubfloat{\label{fig:cond_vs_exp:a}}%
% \phantomsubfloat{\label{fig:cond_vs_exp:b}}%
% \vspace{-2\baselineskip}% Remove extra line inserted by subfloat
\caption{Calculated conductance of a Co atom adsorbed on  Cu(111).
We considered the tip far from the surface, so that the conductance is dominated by the $s$ orbital.
We compare our results with experimental data from  Limot \emph{et al.} \cite{Limot2005} and Knorr \emph{et al.} \cite{Knorr2002}. We present for comparison the conductance calculated without including the surface states in the calculation of the self-energies corresponding to Co on Cu(111) (see \cref{sec:LorentzianG}). 
\label{fig:cond_vs_exp}}
\end{figure}

When the tip is far from the surface, the dominant contribution to the conductance is given by $T_s$ (\cref{eq:Ts}). 
In \cref{fig:cond_vs_exp} we present  $T_s$,  calculated by considering  
$\epsilon_{A1}=0.7$ eV and the bond-pair energy levels splitting. 
In order to infer the importance of a good description of the  surface states,  we also present  the result  obtained  without including the Shockley surface states in the calculation of the self-energies, as discussed in \cref{sec:LorentzianG}.  
We consider  for the conductance calculation a  practically flat $\Gamma^{0\text{-}tip}_{s}$, 
which was adjusted from the result of $\Gamma^{0\text{-}tip}_{s}$ far from the Co atom  (\cref{fig:GammaTip:b}). 
We compare the results with available experimental data, measured by STM \cite{Limot2005,Knorr2002}. 
The experimental curves, given in arbitrary units, are scaled and vertically shifted to facilitate the comparison  with our calculations.

Our results shown in \cref{fig:cond_vs_exp} are in qualitative agreement with the experimental data.
We obtain a dip-like structure, as  observed in the measurements.  
As we pointed out before, the ZBA at the Fermi level is introduced by $G_s^{(c)}$, and is given by the interference of the structure in the Green function of the $d_{z^2}$ orbital with the non-correlated conduction channel given by the $s$ orbital. 
Then, the asymmetry  of the structure presented in \cref{fig:GdGs_zoom_corr:c} is translated into an asymmetric  dip. 
The resonance-like structure at $\omega \approx -0.5$ eV in the  spectra    corresponds to the interaction of the $s$ orbital with the Shockley surface states. This peak was noticed in the calculation without considering multiorbital correlation, \cref{fig:NoCorrelated_vs_DFT_varios}, and is introduced by the independent electron part of the $s$ orbital Green function, $G_{ss}^{0}$ (\cref{eq:Gss0}). 
% % 
In \cref{fig:cond_vs_exp}, we observe that  neglecting the contribution of the surface states in the calculation of Co on Cu(111) completely changes the conductance spectra. 
First, the resonance feature at $\omega \approx -0.5$ eV is absent, since it is directly given by the effect of the surface states on the Anderson widths (see \cref{fig:GammaCu_all:d}). 
In addition, the dip-like structure at the Fermi level is lost, in agreement with the variation observed in \cref{sec:LorentzianG}. 
Both observations show the importance of a proper description of the Cu(111) surface states in the calculation.

\begin{figure}[h!]
\centering
\includegraphics[width=0.95\linewidth,keepaspectratio]{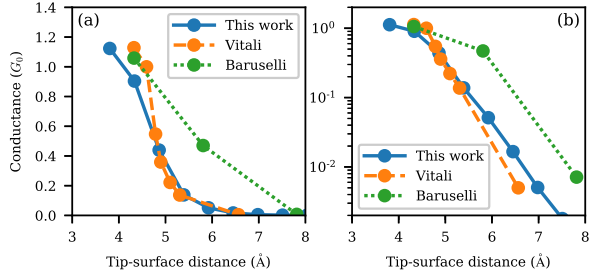}%  
\caption{(a) Calculated conductance for a Co adatom on Cu(111)  as a function of the  tip-surface distance, compared with experimental data from Vitali \emph{et al.} \cite{Vitali2008} and theoretical results from  Baruselli \emph{et al.} \cite{Baruselli2015}. 
Given that the experimental data of Ref. \cite{Vitali2008} is given in terms of relative distances, we rigidly shift the experimental data to better compare with our results. 
This shift is $0.8$ \r{A} smaller than the one used in  Ref. \cite{Baruselli2015} with the same objective.
(b) Shows  the same data in logarithmic scale. 
\label{fig:Current_vs_distance_sinE1_vs_exp_2}}
\end{figure}

In  \cref{fig:Current_vs_distance_sinE1_vs_exp_2} we show the zero-bias conductance  as a function of the  tip-surface distance, compared with available experimental \cite{Vitali2008} and theoretical \cite{Baruselli2015} data.  
Since the experimental distances are relative, we  rigidly shift the experimental curve   along the tip-surface distance axis  to compare with our calculation \cite{Baruselli2015}.
It is worth mentioning that the referenced experimental and theoretical data were compared in Ref. \cite{Baruselli2015}, and  that the shift introduced in the experimental data in Ref. \cite{Baruselli2015} is $0.8$ \r{A} larger than the one used by us in \cref{fig:Current_vs_distance_sinE1_vs_exp_2}. 
The theoretical results of Ref. \cite{Baruselli2015} were calculated using an approach of DFT coupled with numerical renormalization group  \cite{Wilson1974,Bulla2008} (DFT+NRG). 
The authors use the Quantum ESPRESSO DFT code \cite{Giannozzi2009} and PBE functional to obtain the  conductance through  the Co adatom in a supercell approach. 
The results are then used to estimate the parameters for the Anderson impurity model in the wide band limit, by matching the DFT results to the Hartree-Fock   solution of the Anderson model, and the Anderson Hamiltonian is then  solved  using NRG  \cite{Baruselli2015}. 
In  \cref{fig:Current_vs_distance_sinE1_vs_exp_2}, we  observe an overall agreement between our results and the available data.

We stress the importance of a correct description of the surface states  to account for the ZBA line shape (\cref{sec:LorentzianG}). 
In fact, deviations from   experimental  results of theoretical calculations of the ZBA of Co on Cu(111)  have been related  to a poor description of the surface states of Cu(111)  \cite{Baruselli2015,Frank2015}.  
Typical approaches to compute the Hamiltonian parameters, like that used in Ref. \cite{Baruselli2015} or the  embedded cluster calculation used in Ref. \cite{Frank2015} require large supercells in the DFT calculation to describe the surface states \cite{Barral2004} and are usually not considered due to  the associated computational cost. 
Our description of the Hamiltonian parameters allows us to properly describe the coupling of the Co $4s$ orbital with the Cu(111) surface states, leading to an acceptable  agreement with the observed ZBA and also with the measured  tip-surface distance dependence of the conductance.

\section{Conclusions \label{sec:Conclusions}}

We calculated the conductance spectra of a Co atom adsorbed on Cu(111)   from first principles. We considered the Co $3d$ orbitals within a multiorbital correlated model and introduced the $4s$ orbital within a mean-field like approximation. 

Among the $d$ orbitals, only   $d_{z^2}$   couples with the $s$ orbital through the substrate bands. 
The influence of the $d_{z^2}$ orbital in the Co $4s$  spectral density  introduces a zero-bias anomaly in its contribution to the conductance spectra. 
In this way, both conduction channels ($s$ and $d_{z^2}$) interfere to produce a Fano structure in the $s$ contribution to the conductance spectra, which dominates the total conductance when the tip is far from the surface.

Our proposal  satisfactorily describes several features experimentally observed in the conductance spectra  of the Co on Cu(111) system: the dip structure of the zero-bias anomaly around the Fermi energy, the resonance-like structure close to the surface state  low band edge, and the tip-adsorbate distance dependence in the tunneling regime.
Nevertheless, our theoretical  description of the Kondo resonance is limited by our approximated calculation based on the equations of motion method closed up to a second order in the atom-band coupling.

We showed the importance of a proper description of the interaction of the Co adatom with the  surface states present in Cu(111), which can be conceptually extended to other (111) surfaces like Ag(111) and Au(111).
Neglecting the contribution of these states in the Anderson widths completely changes the shape of the zero-bias anomaly, from a dip-like structure to a peak-like structure, and eliminates the resonance feature located close to the onset of the surface state bands.

\begin{acknowledgments}
This work was supported by Consejo Nacional de Investigaciones Cient\'ificas y T\'ecnicas (CONICET) through PIP grants, Universidad Nacional del Litoral (UNL) through CAI+D grants, Deutscher Akademischer Austauschdienst   (DAAD) and Centro Universitario Argentino-Alem\'an (CUAA-DAHZ).
Further support by the German Science Foundation (DFG) through the Collaborative Research Centers SFB-1316 as well as TRR234 is gratefully acknowledged.

\end{acknowledgments}

\appendix

\onecolumngrid
\section{Green functions expressions \label{ap:Derivation}}

We propose the following Green functions, 
\begin{subequations} \label{eq:allG}
 \begin{align}
 \label{eq:Gpq}
G_{pq}(t',t)={}&i\theta(t'-t)\braket{\left\{ \ket{S,M}_{p}\bra{S-\textstyle{\frac{1}{2}},M-\sigma}_{q}(t'),\ket{S-\textstyle{\frac{1}{2}},M-\sigma}_{q}\bra{S,M}_{p}(t)\right\} }
\\ \label{eq:Gss}
G_{ss}(t',t)={}&i\theta(t'-t)\braket{\left\{ \hat{c}_{s\sigma}^{\dagger}(t'),\hat{c}_{s\sigma}(t)\right\} }
\\ \label{eq:Gpqss}
G_{pq}^{ss}(t',t)={}&i\theta(t'-t)\braket{\left\{ \ket{S,M}_{p}\bra{S-\textstyle{\frac{1}{2}},M-\sigma}_{q}(t'),\hat{c}_{s\sigma}(t)\right\} }
\\ \label{eq:Gsspq}
G_{ss}^{pq}(t',t)={}&i\theta(t'-t)\braket{\left\{ \hat{c}_{s\sigma}^{\dagger}(t'),\ket{S-\textstyle{\frac{1}{2}},M-\sigma}_{q}\bra{S,M}_{p}(t)\right\} } .
 \end{align}
\end{subequations}

Since the off-diagonal Anderson widths  of the $d$ orbitals are zero  ($\Gamma_{d_id_j}^{0} = 0$ eV for $i\neq j$), we do not require to compute functions of the form $G_{pq}^{p'q'}$ with $p,q \neq p',q'$. 
We use the equation of motion method to evaluate the evolution of \cref{eq:allG} with the Hamiltonian \eqref{eq:Hionicowiths} and close the system of equations in a second order in the atom-band coupling term. 

In equilibrium, the Fourier transform of  $G_{pq}$ in \cref{eq:Gpq} is given by 
\begin{equation} \label{eq:Gequation}
  G_{pq}(\omega)=\frac{  O_{pq}+ X_{pq}(\omega) }{ \omega-\epsilon_{d(p,q)}- \Sigma_{pq}(\omega)  } .
\end{equation}
The expressions for the terms of \cref{eq:Gequation} are given by 

\begin{subequations} \label{eq:eqdiagwiths}
% \begin{small}
 \begin{align} \label{eq:eqdiagwiths1}
O_{pq} ={}&\braket S_{p}+\braket {S-\textstyle{\frac{1}{2}}}_{q}
\\ \label{eq:eqdiagwiths4}
\Sigma_{pq} ={}&g_{1}\sum_{q'\in p}\Sigma_{d(p,q')}^{>}(\omega-\Delta\epsilon_{d(p,q')}^{d(p,q)})
+g_{S}\sum_{p'\ni q}\Sigma_{d(p',q)}^{<}(\omega-\Delta\epsilon_{d(p',q)}^{d(p,q)})
\\ \nonumber
{}&+g_{1}\sum_{q'\in p}\sigma_{sd(p,q')}(\omega-\Delta\epsilon_{d(p,q')}^{d(p,q)})\Sigma_{sd(p,q')}^{>}(\omega-\Delta\epsilon_{d(p,q')}^{d(p,q)}) 
\\ \nonumber
{}&+g_{S}\sum_{p'\ni q}\sigma_{sd(p',q)}(\omega-\Delta\epsilon_{d(p',q)}^{d(p,q)})\Sigma_{sd(p',q)}^{<}(\omega-\Delta\epsilon_{d(p',q)}^{d(p,q)})
 \\ \label{eq:eqdiagwiths5}
X_{pq} ={}&g_{1}\sum_{q'\in p}\varXi \left[\Sigma_{d(p,q')}^{0},G_{pq'}\right](\omega-\Delta\epsilon_{d(p,q')}^{d(p,q)}) 
-g_{S}\sum_{p'\ni q}\varXi \left[\Sigma_{d(p',q)}^{0},G_{p'q}\right](\omega-\Delta\epsilon_{d(p',q)}^{d(p,q)}) 
\\ \nonumber
&{}+g_{1}\sum_{q'\in p}\varXi \left[\Sigma_{sd(pq')}^{0},G_{pq'}^{ss}\right](\omega-\Delta\epsilon_{d(p,q')}^{d(p,q)})-g_{S}\sum_{p'\ni q}\varXi\left[\Sigma_{sd(p',q)}^{0},G_{p'q}^{ss}\right](\omega-\Delta\epsilon_{d(p',q)}^{d(p,q)})
\\ \nonumber
&{}+g_{1}\sum_{q'\in p}\sigma_{sd(p,q')}(\omega-\Delta\epsilon_{d(p,q')}^{d(p,q)})\left(\varXi\left[\Sigma_{sd(p,q')}^{0},G_{pq'}\right](\omega-\Delta\epsilon_{d(p,q')}^{d(p,q)})+\varXi\left[\Sigma_{s}^{0},G_{pq'}^{ss}\right](\omega-\Delta\epsilon_{d(p,q')}^{d(p,q)})\right)
\\ \nonumber
&{}-g_{S}\sum_{p'\ni q}\sigma_{sd(p',q)}(\omega-\Delta\epsilon_{d(p',q)}^{d(p,q)})\left(\varXi\left[\Sigma_{sd(p',q)}^{0},G_{p'q}\right](\omega-\Delta\epsilon_{d(p',q)}^{d(p,q)})+\varXi\left[\Sigma_{s}^{0},G_{p'q}^{ss}\right](\omega-\Delta\epsilon_{d(p',q)}^{d(p,q)})\right)
.
 \end{align}
%  \end{small}
\end{subequations}

We use the following notation for the occurrence probabilities,
\begin{subequations} \label{eq:occupation}
\begin{align} \label{eq:occupation:p}
\braket{\ket{S,M}_{p}\bra{S,M}_{p}} = {}&  \frac{1}{\pi}\int_{-\infty}^{\infty}d\omega f_{<}(\omega)\text{Im}G_{pq}(\omega) \equiv  \braket S_{p} 
\\ \label{eq:occupation:q}
\braket{\ket{S-\textstyle \frac{1}{2},m}_{q}\bra{S-\textstyle \frac{1}{2},m}_{q }} ={}& \frac{1}{\pi}\int_{-\infty}^{\infty}d\omega f_{>}(\omega)\text{Im}G_{pq}(\omega) \equiv  \braket {S-\textstyle \frac{1}{2}}_{q}  ,
\end{align}
\end{subequations}
which are independent of the spin projection due to the spin degeneracy assumed. The hole occupation of the $d_i$ orbital is obtained from \cref{eq:occupation} by summing over the configurations where the $d_i$ orbital is occupied,
\begin{equation} \label{eq:orbitaloccupation}
o_{d_i}= \gamma_S \sum_{p\ni d_i} \braket S_{p} + \left(\gamma_S-1 \right)\sum_{q\ni d_i} \braket {S-\textstyle \frac{1}{2}}_{q}  ,
\end{equation}
where $\gamma_S=2S+1$ accounts for the sum over the spin projection index.

The total self-energies $\Sigma_{pq}$ are defined in terms of the  Anderson, lesser and greater self-energies,
\begin{equation}
\Sigma_{ab}^{[0/</>]}(\omega)=\sum_{\mathbf{k}}\frac{V_{\mathbf{k}a}^{*}V_{\mathbf{k}b}}{\omega-\epsilon_{\mathbf{k}}-i\eta} f_{[0/</>]}(\epsilon_{\mathbf{k}}) ,
\end{equation}
with $a$ and $b$ replaced by $s$ or $d_i$,   $\Sigma_{a}\equiv \Sigma_{aa}$ and $i \eta$ an  infinitesimal imaginary quantity. The self-energies are evaluated considering the energy level splittings, $\Delta \epsilon_{d(p',q')}^{d(p,q)}= \epsilon_{d(p,q)}-\epsilon_{d(p',q')}$ and coefficients related to  Hund's rule coupling, $g_{1}=  1$  and $g_{S} = 1+\frac{1}{2S}$.
The restriction $q'\in p$ in the sums of \cref{eq:eqdiagwiths} enables only the $q'$ configurations which have allowed transitions with $p$, that is, those for which there is an active orbital $d(p,q')$. A similar restriction, $p' \ni q$, accounts for the complementary case. 
We work   in the hole formalism, where  the Fermi function is defined as 
\begin{equation}
 f_{<}(\epsilon)=1-\frac{1}{1+e^{(\epsilon-\mu)/k_{B}T}}  ,
 \end{equation}
being $\mu$ the chemical potential, $k_{B}$ the Boltzmann constant and $T$ the temperature. We define $ f_{>}(\epsilon)=1-f_{<}(\epsilon)$  and for convenience we introduce the notation $ f_{0}(\epsilon)=1$.

The atom-band derived terms $X_{pq}$ are given by the operator $\varXi$,
\begin{equation} \label{eq:atombandoperator}
\varXi \left[\Sigma,G \right] (\tilde{\omega})=\frac{1}{\pi}\int_{-\infty}^{\infty} d\omega' \frac{f_{<}(\omega')}{\tilde{\omega}-\omega'-i\eta}\left(\text{Im}\left(\Sigma(\omega')G(\omega')\right)-\Sigma(\tilde{\omega})\text{Im}G(\omega')\right) .
\end{equation}

The effects  introduced by the $s$ orbital on the $d$ orbitals  are related to the factor
\begin{equation}
 \sigma_{sd_{i}}(\omega)=\frac{\Sigma_{sd_{i}}^{0}(\omega)}{\omega-\epsilon_{s}-\Sigma_{s}^{0}(\omega)} ,
\end{equation}
which involves the off-diagonal self-energy $\Sigma_{sd_{i}}^{0}$ between the $s$   and   $d_i$ orbitals. 

The Green function of  \cref{eq:Gpqss},  involving operators of   $s$ and  $d$ orbitals, is given by
\begin{align}
\label{eq:Gpqsseq}
 G_{pq}^{ss}= \sigma_{sd(p,q)}(\omega)G_{pq}(\omega)
\end{align}
and a similar expression is found for \cref{eq:Gsspq}. 

The Green function related to the $s$ orbital is given by 
\begin{equation} \label{eq:Gsseq}
 G_{ss}(\omega)=G_{ss}^{0}(\omega)+\frac{\gamma_{S}}{2}\sum_{p q} \left( \sigma_{sd(p,q)}(\omega) \right)^2   G_{pq}(\omega)   .
\end{equation}
The independent particle Green function $G_{ss}^{0}$ is 
\begin{equation} \label{eq:Gss0ap}
 G_{ss}^{0}(\omega)=\frac{1}{\omega-\epsilon_{s}-\Sigma_{s}^{0}(\omega)} .
\end{equation}
The occupation of the $s$ orbital is given by
\begin{equation} \label{eq:occs}
 O_{ss} = \frac{1}{\pi}\int_{-\infty}^{\infty} d\omega f_{<}(\omega) \text{Im}G_{ss}(\omega) .
\end{equation}

\Cref{eq:Gequation,eq:Gsseq} are then used to define the orbital Green functions of \cref{eq:allGds}, where the  sum  over the spin index is done.

\section{Configurations \label{ap:configurations}}

The possible configurations of Co on  Cu(111) are   presented in Tables \ref{tab:PQfluctuations:a} and \ref{tab:PQfluctuations:b}. 
They correspond to the five non-equivalent possibilities in which the orbitals of the states with $S=\frac{3}{2}$ ($P$) can be filled with three holes, and the five possibilities to fill the $d$ orbitals with two holes in order to build the $ S-\frac{1}{2}=1$ ($Q$) configurations.
In the same way, the $11$ non-equivalent fluctuations  giving place to the required Green functions are those of \cref{tab:PQfluctuations:c}. 

\begin{table}[h]
\begin{center}
\subfloat[$S=\frac{3}{2}$. \label{tab:PQfluctuations:a}]{
\begin{tabular}{cc}
\toprule
$P$  \\  
\midrule 
$E1 E1 E2$  \\
$E1 E2 E2$  \\
$E1 E1 A1$   \\
$E1 E2 A1$   \\
$E2 E2 A1$   \\
\bottomrule
\end{tabular}
}\quad \quad
\subfloat[$S-\frac{1}{2} $. \label{tab:PQfluctuations:b}]{
\begin{tabular}{ccc}
\toprule
&  $Q$ & \\  
\midrule 
& $E1 E1$ &   \\
& $E1 E2$ &  \\
& $E2 E2$ & \\
& $E1 A1$ &  \\
& $E2 A1$ & \\
\bottomrule
\end{tabular}
}\quad \quad
\subfloat[Non-equivalent fluctuations. \label{tab:PQfluctuations:c}]{
\begin{tabular}{cccccccc}
\toprule
Fluctuation & $P$         &  $Q$      & $D(P,Q)$ &  Fluctuation &  $P$      &    $Q$     &  $D(P,Q)$  \\
\midrule                                                                                                                                   
 1          & $E1 E1 E2$  &  $E1 E1$  &   $E2$   &     7       & $E1 E2 A1$ &   $E1 A1$  &    $E2$     \\
 2          & $E1 E1 E2$  &  $E1 E2$  &   $E1$   &     8       & $E1 E2 A1$ &   $E1 E2$  &    $A1$           \\
 3          & $E1 E2 E2$  &  $E1 E2$  &   $E2$   &     9       & $E1 E2 A1$ &   $E2 A1$  &    $E1$           \\
 4          & $E1 E2 E2$  &  $E2 E2$  &   $E1$   &     10      & $E2 E2 A1$ &   $E2 A1$  &    $E2$         \\
 5          & $E1 E1 A1$  &  $E1 E1$  &   $A1$   &     11      & $E2 E2 A1$ &   $E2 E2$  &    $A1$       \\
 6          & $E1 E1 A1$ &   $E1 A1$  &   $E1$   &             &             &       &       \\
\bottomrule
\end{tabular}
}
\end{center}
\caption{The five non-equivalent possibilities for (a) $S=\frac{3}{2}$ ($P$) and (b) $S-\frac{1}{2}=1$ ($Q$), to accommodate the corresponding holes into the three symmetry groups: $E1$, $E2$ and $A1$.
(c) Non-equivalent fluctuations between the five  sets of configurations with three holes ($P$) and the five with two holes ($Q$) of the Co orbitals splitted into the three symmetry groups. The symmetry of the active orbital involved in  the transition is indicated by $D(P,Q)$.
}
\label{tab:PQfluctuations}
\end{table}

\twocolumngrid

% \bibliography{Main}


\end{document}